\documentclass[sigconf]{acmart}
\usepackage{booktabs}
\usepackage[table]{xcolor}
\usepackage{multirow}
\usepackage{color,soul}
\usepackage{algorithm}
\usepackage{algorithmic}
\usepackage{arydshln}

\usepackage{threeparttable}
\usepackage{hyperref, listings, enumitem}
\usepackage{multirow}

\definecolor{myorange}{HTML}{FF7F0E}
\definecolor{myblue}{HTML}{1F77B4}
\AtBeginDocument{%
  }

\setcopyright{acmlicensed}
\copyrightyear{2026}
\acmYear{2026}
\acmConference[CCS '26]{ACM Conference on Computer and Communications Security}{November 15-19, 2026}{The Hague, The Netherlands}

\begin{document}

\title{Privacy Enhanced PEFT:
Tensor Train Decomposition Improves Privacy Utility
Tradeoffs under DP-SGD}

\author{Pradip Kunwar}

\orcid{0009-0004-2583-5925}
\affiliation{%
  \institution{Tennessee Tech University, \\Los Alamos National Laboratory}
  \city{Cookeville}
 \state{Tennessee}
 \country{USA}
}
\email{pkunwar42@tntech.edu}

\author{Minh N. Vu}
\orcid{0000-0001-8727-0350}
\affiliation{%
  \institution{Los Alamos National Laboratory}
  \city{Los Alamos}
 \state{New Mexico}
 \country{USA}
  }
\email{mvu@lanl.gov}

\author{Maanak Gupta}
\orcid{0000-0001-9189-2478}
\affiliation{%
  \institution{Tennessee Tech University}
   \city{Cookeville}
 \state{Tennessee}
 \country{USA}
}
\email{mgupta@tntech.edu}

\author{Manish Bhattarai}
\orcid{0000-0002-1421-3643}
\affiliation{%
  \institution{Los Alamos National Laboratory}
  \city{Los Alamos}
 \state{New Mexico}
 \country{USA}
  }
\email{ceodspspectrum@lanl.gov}

\renewcommand{\shortauthors}{Kunwar et al.}

%%% ABSTRACT %%%
\begin{abstract}

Fine-tuning large language models on sensitive data poses significant privacy risks, as membership inference attacks can reveal whether individual records were used during training. While Differential Privacy (DP) provides formal protection, applying DP to conventional Parameter-Efficient Fine-Tuning (PEFT) methods such as Low-Rank Adaptation (LoRA) often incurs substantial utility loss. In this work, we show that a \emph{more structurally constrained} PEFT architecture, Tensor Train Low-Rank Adaptation (TTLoRA), can improve the privacy–utility tradeoff by shrinking the effective parameter space while preserving expressivity. To this end, we develop TTLoRA-DP, a differentially private training framework for TTLoRA. Specifically, we extend the ghost clipping algorithm to Tensor Train cores via cached contraction states, enabling efficient Differentially Private Stochastic Gradient Descent (DP-SGD) with exact per-example gradient norm computation without materializing full per-example gradients. Experiments on GPT-2 fine-tuning over the Enron and Penn Treebank datasets show that TTLoRA-DP consistently strengthens privacy protection relative to  LoRA-DP while maintaining comparable or better  downstream utility. Moreover, TTLoRA exhibits lower membership leakage even without DP training, using substantially smaller adapters and  requiring on average $7.6\times$ fewer parameters than LoRA. Overall, our results demonstrate that TTLoRA offers a practical path to improving the privacy–utility tradeoff in parameter-efficient language model adaptation.

%To this end, we introduce {TTLoRA-DP}, the first differentially private training framework for Tensor Train Low-Rank Adaptation (TTLoRA). 

%To enable efficient \hl{DP-SGD}, we extend ghost clipping to Tensor Train (TT) cores via cached contraction states, allowing exact per-example gradient norm computation without materializing full per-example gradients. 
%Experiments on GPT-2 fine-tuning over the Enron and Penn Treebank datasets show that TTLoRA-DP consistently achieves stronger privacy protection than \hl{DP-LoRA} while achieving more competitive downstream utility. Moreover, TTLoRA exhibits lower membership leakage even without DP training with substantially smaller adapters, requiring on average $7.6\times$ fewer parameters than LoRA. \hl{Overall, our results demonstrate that tensor train decomposition provides an effective structural inductive bias for privacy-preserving language model adaptation.}

\end{abstract}

% Our key insight is that TTLoRA's hierarchical tensor decomposition naturally amplifies the privacy benefits.





\maketitle

\section{Introduction}
The deployment of Large Language Models (LLMs) in privacy-sensitive settings—such as healthcare, finance, and legal services often requires fine-tuning on confidential corpora \cite{brown2022does, carlini2021extracting}. While Full Fine-Tuning (FFT) achieves high utility, its massive parameter count poses significant challenge for computational efficiency \cite{houlsby2019parameter, hu2022lora}. Parameter -Efficient Fine-Tuning (PEFT) methods, most notably LoRA, address the challenge by adapting only a small subset of parameters, significantly reducing memory and computation requirements \cite{hu2022lora}. However, LoRA’s two-factor matrix design imposes a fundamental \emph{parameter floor}, restricting how lightweight the adapters can be without degrading performance. Tensor Train Low-Rank Adaptation (TTLoRA) \cite{anjum2024tensor}, a structurally constrained PEFT that leverages Tensor Train (TT) decomposition \cite{oseledets2011tensor, novikov2015tensorizing}, overcomes this limitation. It organizes the adaptation into a 3rd-order chain of tensor cores, enabling an ultra-low-parameter regime that maintains expressivity and competitive task performance \cite{kunwar2025tt}.

% Parameter-Efficient Fine-Tuning (PEFT) methods, most notably LoRA, have emerged as the efficient and standard approach for adapting models using only a fraction of the parameters \cite{hu2022lora}. However, LoRA's two-factor matrix structure introduces a fundamental \emph{parameter floor}, as each adaptation layer requires $r(d_{in}+d_{out})$ parameters. For a model like GPT-2, even an ultra-low rank 2, LoRA requires $\sim$110.6K trainable parameters, and further reducing rank restricts representational capacity, degrading utility \cite{hu2022lora, biderman2024lora, rathore2025much}. TTLoRA overcomes the \emph{parameter floor} bottleneck by leveraging Tensor Train Decomposition structure \cite{oseledets2011tensor, novikov2015tensorizing}, enabling adaptation in an ultra-low parameter regime that achieves a $14.5\times$ reduction over rank-2 LoRA while maintaining competitive performance across diverse tasks \cite{anjum2024tensor, kunwar2025tt}. 
% Such extreme parameter efficiency is particularly attractive for privacy-preserving deployment, as the effectiveness of mechanisms like Differential Privacy (DP) \cite{dwork2006calibrating} is inversely proportional to the number of trainable parameters \cite{bassily2014private, abadi2016deep, bun2014fingerprinting}. 

Beyond parameter efficiency, how PEFT architectures shape model privacy leakage remains underexplored. Fine-tuning on sensitive datasets raises risks of Membership Inference Attacks (MIA) \cite{shokri2017membership, zarifzadeh2023low, liu2024precurious}. While LoRA exhibits some inherent privacy benefits—due to implicit regularization and noise-like optimization dynamics \cite{malekmohammadi2024low, das2025revisiting}—these effects are often insufficient, leaving models vulnerable \cite{ran2025lora,liu2024precurious}. Differential Privacy (DP) can mitigate the leakage \cite{ran2025lora, ma2024efficient}, but applying DP to LoRA typically comes at a substantial utility cost \cite{xu2024dp, sun2024improving, das2025revisiting}. This raises a key question: \emph{Can a more structurally constrained PEFT method like TTLoRA yield stronger inherent privacy and more favorable DP utility trade-offs while operating in an ultra-low parameter regime?}

% Beyond parameter count, the impact of PEFT architectures, particularly TTLoRA, on model privacy remains unexplored. As fine-tuning on sensitive datasets becomes increasingly ubiquitous, concerns about privacy leakage have intensified \cite{shokri2017membership, carlini2019secret, carlini2021extracting, zarifzadeh2023low, liu2024precurious}, making it essential to understand how adaptation structure influences privacy behavior. Recent studies suggest that LoRA’s optimization dynamics demonstrate inherent privacy benefits, resembling implicit noise injection or regularization that suppresses member-specific overfitting \cite{malekmohammadi2024low, das2025revisiting}. However, these intrinsic effects are often insufficient in practice, as LoRA remains highly vulnerable to Membership Inference Attacks (MIA) \cite{ran2025lora, liu2024precurious}. One common approach toward defense is to apply Differential Privacy (DP) to mitigate the leakage \cite{ran2025lora, ma2024efficient}; however, this typically results in substantial utility degradation \cite{xu2024dp, sun2024improving, das2025revisiting}. Together with these observations, we raise a fundamental question: \emph{Can a more structurally constrained tensorized PEFT method, TTLoRA, enable stronger inherent privacy and improve the privacy–utility trade-offs under DP while maintaining ultra-low parameter counts?}
 In this work, we show that TTLoRA empirically delivers a more favorable privacy–utility tradeoff than LoRA, both with and without DP training. The tensor train structure enables adapters that are simultaneously smaller, expressive \cite{kunwar2025tt}, and implicitly regularized \cite{razin2021implicit}, reducing sample-specific memorization. Building on this, we develop \textbf{TTLoRA-DP}, the first Differentially Private Stochastic Gradient Descent (DP-SGD) pipeline for TTLoRA, and demonstrate that it consistently improves empirical privacy (MIA robustness) and retains competitive utility under both DP and non-DP training.

% We show that the \emph{\textbf{answer is yes}}: tensor train structure enables smaller adapters that simultaneously improve empirical privacy protection and preserve utility in both DP-SGD and without DP-SGD. We adopt Tensor Train (TT) factorization \cite{oseledets2011tensor, novikov2015tensorizing}, which decomposes weight updates ($\Delta W$) into a sequence of tensor cores, enabling smaller, expressive adapters \cite{kunwar2025tt} with implicit regularization \cite{razin2021implicit}. We implement \textbf{TTLoRA-DP}, the first DP-SGD pipeline for TT adapters, enabling efficient privacy-preserving training. 
% To realize these benefits under DP-SGD, we develop novel \textbf{TTLoRA-DP} by extending ghost clipping \cite{li2021large} to TT-cores via cached contraction states, enabling per-example gradient norm computation without materializing full per-example gradient tensors.

\begin{figure}[t]
    \centering
    \includegraphics[width=\linewidth]{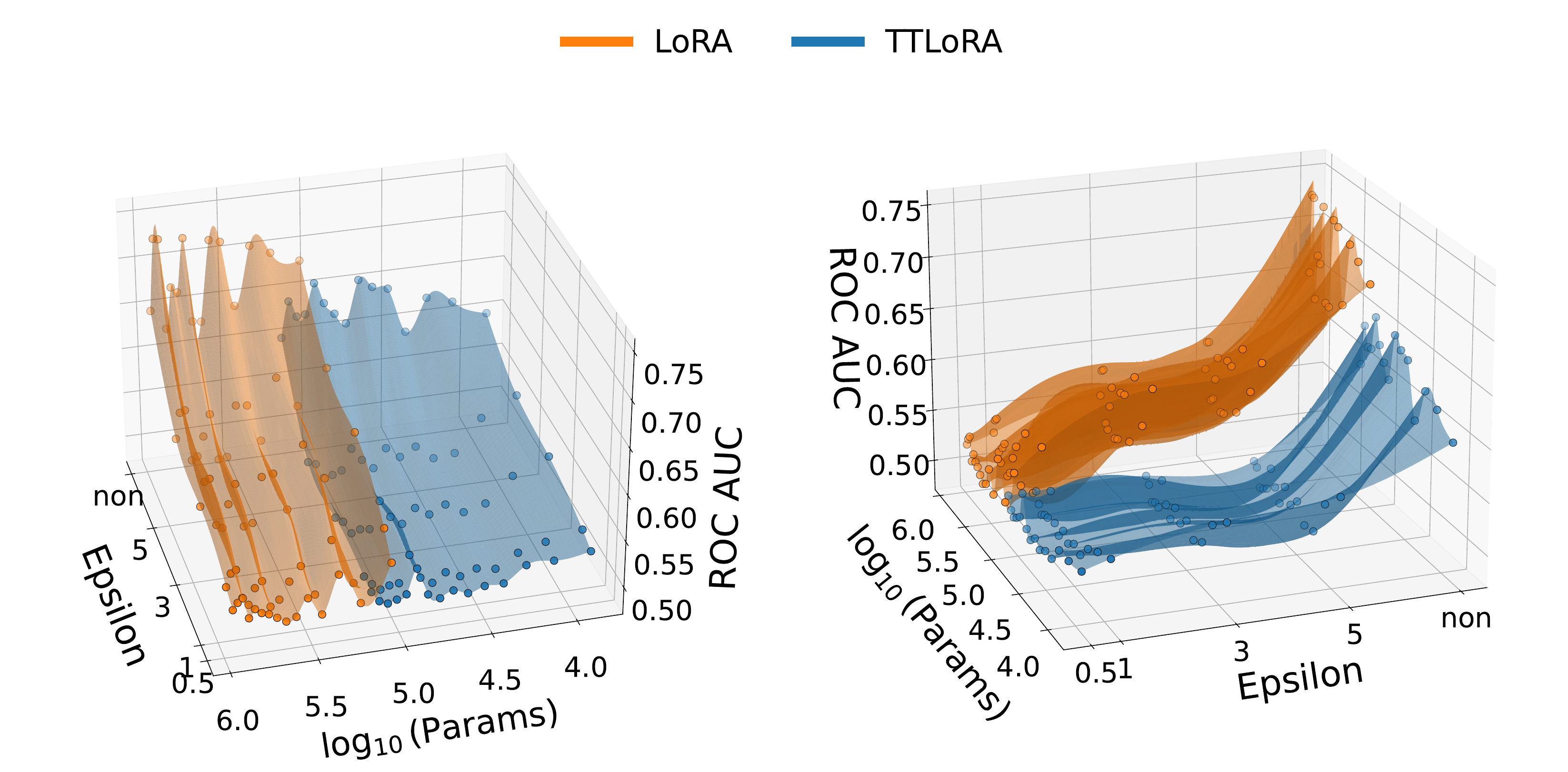}
    \caption{Attack AUC 3D surface on Enron dataset (two views). LoRA shows significant variation in attack AUC (lower is better) across epsilon whereas TTLoRA maintains a relatively flat surface near the 50\% random-guess baseline in private setting (non here means non-private setting).}
    \label{fig:3dplot}
\end{figure}

\begin{figure}[h]
  \centering
  \includegraphics[width=\linewidth]{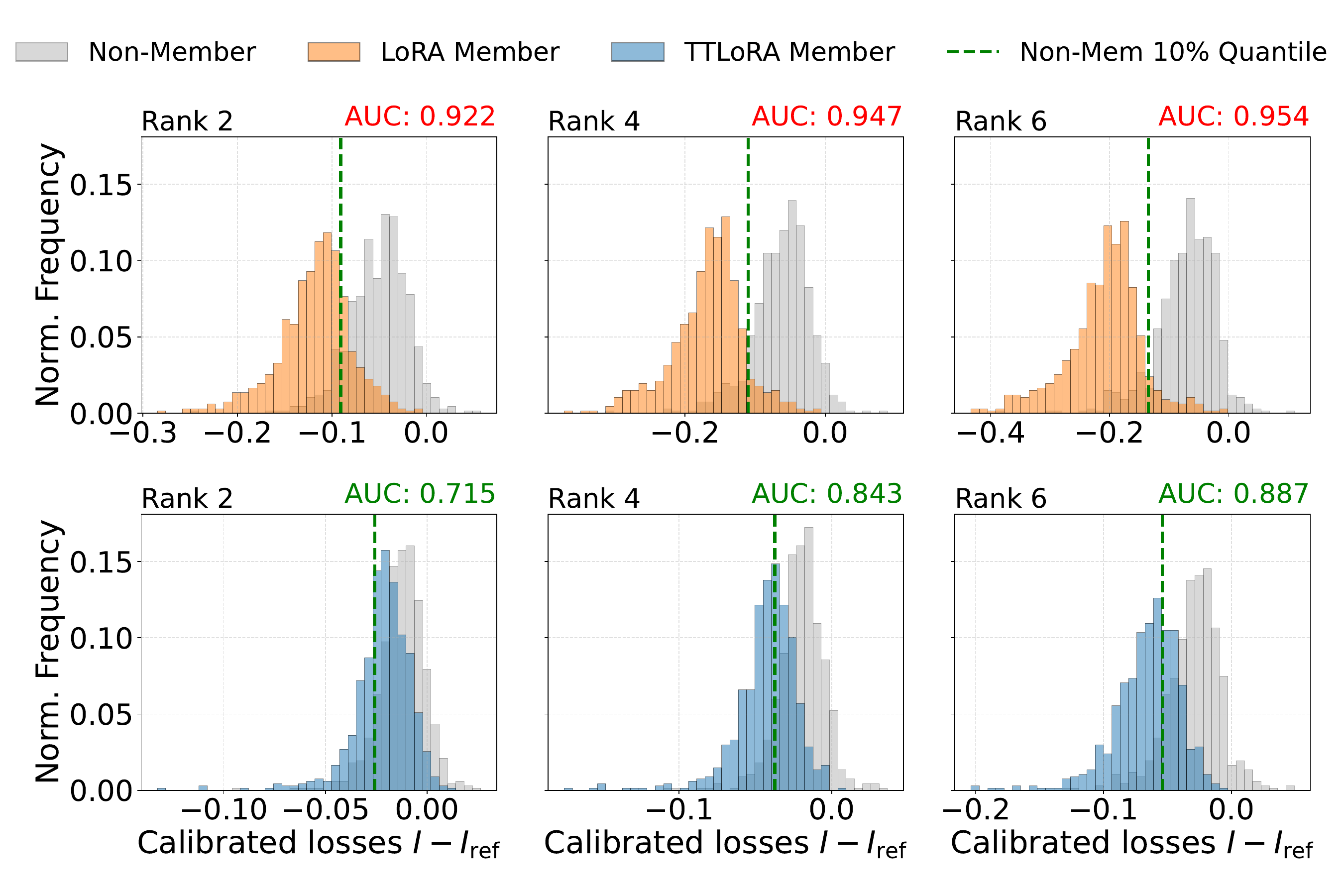}
  \caption{\textbf{Calibrated loss distributions for MIA leakage (Enron) under non-private setting.} Histograms show $\,\mathcal{L}(x;\theta_{\mathrm{peft}})-\mathcal{L}(x;\theta_{\mathrm{ref}})\,$ for members and non-members at ranks 2/4/6 for LoRA and TTLoRA; dashed line marks the 10\% non-member quantile used for fixed-FPR evaluation. Larger separation implies higher AUROC and stronger leakage.}
  \label{fig:loss-distribution}
\end{figure}

We evaluate the impact of TTLoRA’s tensor train structure along two axes that are critical in practice: (i) \textbf{DP utility}—measuring task performance across different privacy budgets $\varepsilon$ (smaller $\varepsilon$ indicates stronger privacy), and (ii) \textbf{empirical leakage under DP}—tracking performance versus membership leakage MIA AUC (lower is better) across the same privacy budgets. We further investigate whether the observed effects persist \textbf{without DP}, measuring membership leakage against performance. To quantify leakage, we implement the state-of-the-art loss-based attack protocol PreCurious\footnote{https://github.com/Emory-AIMS/PreCurious} ~\cite{liu2024precurious}. For DP evaluation, we consider multiple privacy budgets, $\varepsilon = \{0.5, 1, 3, 5\}$, using RDP accounting \cite{mironov2017renyi} with $\delta = 8 \times 10^{-6}$ or $N^{1.1}$ (where $N$ = number of samples). Specifically, our contributions are as follows:

% We evaluate TT architectural hypothesis along the two axes that matter in practice—DP utility (performance vs. $\varepsilon$), empirical leakage (performance vs. MIA AUC vs. $\varepsilon$),—and additionally test whether the effect persists without DP (performance vs MIA AUC). To measure the membership leakage, we implement the state-of-the-art loss based attack protocol (PreCurious)~\cite{liu2024precurious}, yielding higher attack success rate. Moreover for DP implementation, we choose multiple $\varepsilon=\{0.5,1,3,5\}$ budgets accounted with RDP \cite{mironov2017renyi} accounting with $\delta=8e^{-6}$ or $N^{1.1}$ (where N is the number of data samples).
% Our work demonstrates that TTLoRA’s tensor train structure not only reduces effective dimensionality but also reshapes optimization dynamics to improve both inherent and DP-enabled privacy. Specifically, \textbf{our contributions} and key findings are as follows:
\begin{itemize}
    \item \textbf{Higher Utility under Private Training}: TTLoRA consistently achieves stronger utility under DP compared to LoRA across all $\varepsilon$ budgets. For example, on the Enron dataset at a strict privacy budget of $\varepsilon=0.5$, TTLoRA achieves an average perplexity of 27.52 (lower is better), outperforming LoRA’s 29.29 (see Table \ref{tab:dp-ppl}).
    \item \textbf{Robustness against MIA Leakage under DP}: When tested under PreCurious attack protocol, TTLoRA fundamentally alters how membership leakage scales with the privacy budget. On Enron, as $\varepsilon$ increases, LoRA exhibits a significant rise in vulnerability, with average attack AUC (MIA leakage) increasing from 52.49\% to 58.43\%, whereas TTLoRA remains nearly stable, changing only from 51.36\% to 52.04\% (see Fig. \ref{fig:3dplot} and Table \ref{tab:dp-mia-avg-comparison}). Importantly, these privacy gains incur negligible utility loss, with perplexity differences consistently limited to 0.01.
    \item \textbf{Stronger Inherent Privacy Without DP}: Beyond DP, TTLoRA exhibits substantially lower membership leakage even without DP-SGD. On Enron, TTLoRA reduces average attack AUC (MIA leakage) to 88.68\% and FPR@1\% to 16.19\%, compared to 94.91\% and 33.47\% for LoRA, while maintaining comparable perplexity (18.05 vs. 17.46). Calibrated loss distributions (Fig. \ref{fig:loss-distribution}) show significantly greater overlap between member and non-member losses, indicating reduced sample-specific memorization due to the tensor train structure.
\end{itemize}
Taken together, these results indicate that privacy-preserving adaptation of LLMs need not rely solely on stronger DP mechanisms or tighter privacy budgets. Instead, the structured tensor train decomposition in TTLoRA—which imposes a compositional constraint while reducing the effective parameter space—significantly shapes both utility and privacy leakage. By operating well below the practical parameter floor of conventional LoRA, TTLoRA enables effective private fine-tuning with ultra-low-parameter adapters.

%\section{Organization}
%\label{sec:organization}
The remainder of this paper is organized as follows. Section~\ref{sec:relatedwork} provides overview of the related work. In Section~\ref{sec:prelim}, we establish the necessary preliminaries by reviewing Tensor Train (TT) decomposition, detailing how TTLoRA employs TT factorization to construct parameter-efficient adapters, and the challenge of per-example gradient computation of TTLoRA in Differential Privacy. In Section \ref{sec:ghost-clipping}, we introduce TTLoRA-DP framework; here, we describe our primary technical contribution: an extension of the ghost clipping algorithm to TT-cores. Section \ref{sec:analysis} provides a rigorous empirical analysis of TTLoRA-DP, evaluating its resilience against membership inference attacks and its utility-privacy trade-offs. Finally, Section \ref{sec:discussion} and \ref{sec:conclusion} highlights the discussion of broader impacts and concluding remarks.

\section{Related Work}
\label{sec:relatedwork}

\paragraph{\textbf{Parameter-Efficient Fine-Tuning (PEFT)}}
PEFT methods adapt large pretrained models by updating only a small subset of parameters, reducing training and deployment costs. Among these methods \cite{houlsby2019parameter,li2021prefix, lester2021power}, LoRA has become the widely adopted standard due to its simplicity, factorizing weight updates into two low-rank matrices \cite{hu2022lora}. However, LoRA imposes a fundamental \emph{parameter floor}: the two-factor matrix structure enforces a minimum parameter count per adapted layer, limiting scalability in ultra-low-parameter regimes.

This has motivated \emph{structured} PEFT parameterizations that further constrain the update space while retaining expressivity. Tensor-based approaches address this limitation by introducing stronger structural constraints. LoRETTA uses tensor-train (TT) decomposition \cite{oseledets2011tensor} to reduce trainable parameters beyond conventional low-rank designs \cite{yang2024loretta}. Closely related, TTLoRA and follow-up work explore TT-based parameterizations for fast, compressed fine-tuning and extensions to sparse MoE variants \cite{anjum2024tensor,kunwar2025tt}. Prior works establish tensor train's parameter compression capabilities but do not study its privacy implications. Our work focuses on an underexplored dimension: how TT structure (using TTLoRA), impacts \emph{privacy leakage} and the \emph{privacy--utility} frontier.

\paragraph{\textbf{Privacy Leakage and Membership Inference.}}
Fine-tuned language models are vulnerable to membership inference attacks (MIA), which infer whether a data point was used during training \cite{shokri2017membership,yeom2018privacy}. A particularly relevant recent direction identifies new fine-tuning attack surfaces that exploit the \emph{pretrained checkpoint}. \emph{PreCurious} shows that an adversary can publish a crafted pretrained model that later amplifies privacy leakage during a victim’s fine-tuning, enabling strong black-box MIAs via calibrated losses \cite{liu2024precurious}. Relatedly, \emph{Privacy Backdoors} demonstrates that poisoning pretrained models can significantly increase membership leakage after fine-tuning, underscoring the risks of untrusted initialization checkpoints \cite{wen2024privacybackdoors}. \emph{Robust membership inference attack (RMIA)} models the null hypothesis in likelihood-ratio testing and leverages both reference models and population data to achieve high test power across the TPR–FPR curve  \cite{zarifzadeh2023low}. Besides, our work studies PEFT privacy leakage and DP defenses in a fine-tuning pipeline, PreCurious provides a strong and practically motivated threat model for our evaluation.

\paragraph{\textbf{Differential Privacy for Fine-Tuning LLM}}
Differential privacy (DP) \cite{dwork2006calibrating} limits how much any single training example can influence a model’s outputs, and in deep learning this is typically enforced by DP-SGD \cite{abadi2016deep}. Despite its formal guarantees, the standard DP-SGD mechanism remains difficult to deploy for LLM. DP-SGD requires per-example gradient norm computation for clipping, which is memory and compute-intensive at LLM scale and has motivated specialized implementations such as ghost clipping and fast per-example clipping \cite{abadi2016deep, goodfellow2015efficient, lee2021scaling, li2021large, bu2022scalable}. 

Even when training is made computationally feasible, the challenges arise from the fact that the magnitude of DP noise scales with the dimensionality of the trainable parameter space \cite{bassily2014private, abadi2016deep, bun2014fingerprinting}. Reducing the number of parameters involved in training therefore directly improves both computational overhead and noise-induced utility degradation \cite{yu2021differentially, ma2024efficient}. PEFT methods address this issue by restricting learning to a small subset of parameters, making them a natural substrate for private fine-tuning. 

\paragraph{\textbf{Differential Privacy Meets PEFT}}
Because PEFT restricts the trainable parameter space, this raises the question of whether PEFT \emph{reduces privacy leakage} and whether it can improve DP utility. Recent works provide mixed but increasingly nuanced answers. On one hand, theoretical and empirical studies suggest that LoRA inherently acts like it injects noise-like effects into optimization and may partially resemble DP-SGD dynamics, with privacy behavior depending on rank and model dimensions \cite{malekmohammadi2024low, das2025revisiting}. On the other hand, recent attacks specifically targeting LoRA fine-tuning show that LoRA-adapted models can remain largely vulnerable to MIAs \cite{ran2025lora}, especially when the attacker leverages the publicly available pretrained model as a reference \cite{liu2024precurious}.

Several works explicitly study DP with PEFT and/or federated learning, proposing private low-rank adaptation mechanisms or improving LoRA-based federated training under privacy constraints \cite{sun2024improving,xu2024dp,liu2025differentially}. Closest to our motivation, Ma et al.\ investigate memorization under DP parameter-efficient fine-tuning and show that PEFT can reduce leakage while retaining performance, but the role of \emph{PEFT structure} is still not fully characterized \cite{ma2024efficient}. In contrast to prior DP-PEFT studies that primarily focus on conventional LoRA, our work investigates whether \emph{more structurally constrained} TT parameterized TTLoRA architecture can deliver stronger privacy--utility behavior and provides the first DP-SGD pipeline for TTLoRA.

% \subsection{Positioning of Our Work}
% Existing ghost clipping implementations support standard layers and LoRA-style adapters \cite{li2021large}, but do not natively support TTLoRA’s chained TT-core parameterization. This leaves a gap between TT-structured PEFT methods (motivated by extreme parameter efficiency) and practical DP-SGD training infrastructure. Our contribution bridges this gap by (i) extending ghost clipping to TT core chains with cached contraction states to enable exact per-example gradient norm computation efficiently, and (ii) empirically demonstrating that TTLoRA yields a more favorable privacy--utility profile than LoRA under both DP and non-DP training, including under the strong PreCurious threat model \cite{liu2024precurious}.

\section{Preliminaries}
\label{sec:prelim}
\subsection{Tensor Train Decomposition}
\label{prelim:ttd}
The Tensor Train (TT) decomposition \cite{oseledets2011tensor, novikov2015tensorizing} compresses  a high order $n$-dimensional tensor $\mathcal{T} \in \mathbb{R}^{I_1 \times I_2 \times \cdots \times I_n}$ by factorizing it into a sequence of 3rd-order  tensors called TT-cores. Specifically, each element of $\mathcal{T}$ is computed by multiplying the corresponding slices of these cores as:
\begin{equation}
    \mathcal{T}(i_1, i_2, \ldots, i_n) = G_1[:, i_1, :] \cdot G_2[:, i_2, :] \cdots G_n[:, i_n, :]
\end{equation}
where $G_k \in \mathbb{R}^{r_{k-1} \times I_k \times r_k}$ are the TT-cores and ${r_0, r_1, \dots, r_n}$ are the TT-ranks (with  $r_0 = r_n = 1$). This reduces storage complexity from $\mathcal{O}(\prod I_k)$ to $\mathcal{O}(\sum r_{k-1} I_k r_k)$, enabling extremely compact parameterization. %Intuitively, TT decomposition compresses the original tensor: instead of storing $\prod_k I_k$ elements, we only need $\sum_k r_{k-1} I_k r_k$ elements. This makes TT decomposition particularly attractive for parameter-efficient adaptations like TTLoRA, where large weight updates can be represented compactly while retaining expressivity.

\subsection{TTLoRA Architecture}
\label{subsec:ttloraarch}
%Building on the TT decomposition as described in Section \ref{prelim:ttd},
TTLoRA is fine-tuning method based on TT decomposition. TTLoRA applies TT decomposition directly to the weight update $\Delta W \in \mathbb{R}^{d_\text{out} \times d_\text{in}}$. By representing $\Delta W$ as a chain of tensor cores, it  achieves two key goals: (i) drastically reduces the number of trainable parameters, thereby breaking the LoRA parameter floor \cite{kunwar2025tt}, and (ii) introduces a structured compositional constraint that implicitly regularizes the adaptation \cite{razin2021implicit}, which we hypothesize contributes to improved privacy and robustness.

\paragraph{\textbf{Factorization}} The weight update $\Delta W \in \mathbb{R}^{d_\text{out} \times d_\text{in}}$ is factorized along both input and output dimensions as: 
\begin{equation}
d_\text{in} = \prod_{k=1}^{p} m_k, \qquad d_\text{out} = \prod_{\ell=1}^{q} n_\ell
\end{equation}
The adapter is then represented using $(p+q)$ TT-cores: input cores $G_k \in \mathbb{R}^{r_{k-1} \times m_k \times r_k}$ and output cores $\widetilde{G}\ell \in \mathbb{R}^{r_{p+\ell-1} \times n_\ell \times r_{p+\ell}}$, with boundary ranks $r_0 = r_{p+q} = 1$ and internal ranks typically set uniformly to $r$. To illustrate, consider GPT-2’s attention weight of size $768 (d_{out}) \times 2304(d_{in})$, which can be factorized into $[64, 4, 3] \times [3, 3, 4, 64]$. With TT-rank $r=2$, the output TT-cores can be represented as $[1,64,2]$, $[2,4,2]$, and $[2,3,2]$ whereas the input cores as $[2,3,2]$, $[2,3,2]$, $[2,4,2]$, and $[2,64,1]$. In practice, the key hyperparameters controlling TTLoRA are the TT-rank $r$, the chosen factorization (TT-shape) of input and output dimensions, and the scaling factor $\alpha$. During the forward pass, the input activation tensor is sequentially contracted with the input cores (compressing information into the rank dimension) and then expanded through the output cores. More details about tensor operations can be found in Section \ref{sec:notation}--\ref{sec:expandingoutput}.

\subsection{Differential Privacy and Challenges}
Differential Privacy (DP) \cite{dwork2006calibrating} provides a rigorous framework to bound privacy leakage from training data. A randomized mechanism $\mathcal{M}$ satisfies $(\varepsilon, \delta)$-DP if, for any two adjacent datasets $D$ and $D'$ differing in a single record, and any measurable set $S$:
\begin{equation}
    \Pr[\mathcal{M}(D) \in S] \leq e^\varepsilon \Pr[\mathcal{M}(D') \in S] + \delta
\end{equation}
While DP offers formal privacy guarantees, applying it to LLM fine-tuning is challenging. Standard DP-SGD requires three key steps: (i) computing per-example gradient norms, (ii) clipping each gradient to a fixed norm to bound sensitivity, and (iii) adding calibrated noise to the aggregated gradients before updating the model. Computing exact per-example gradient norms for clipping is memory and compute intensive at LLM scale, especially for structured adapters.
 Ghost clipping \cite{li2021large} addresses the computational bottleneck by avoiding materialization of full per-example gradients, enabling efficient DP-SGD. 

\textbf{The Challenge for TTLoRA:} Existing ghost clipping implementations\footnote{https://github.com/lxuechen/private-transformers} support standard layers (e.g., linear, convolutional, embeddings) and LoRA-style adaptation, but do not handle TTLoRA’s chained TT-core parameterization. In TTLoRA, the weight update is represented implicitly as a product of small TT-cores and applied to an input via sequential core contractions. As a result, applying standard ghost clipping would require explicitly forming $\Delta W$ (or full per-example gradients), which is memory-prohibitive and negates the efficiency benefits of PEFT. The core mismatch is therefore structural: the per-example Jacobians needed for clipping are distributed across the TT-cores rather than associated with a single matrix parameter. To resolve this, we extend ghost clipping to TTLoRA by performing sequential contraction/expansion with cached states, enabling exact per-example gradient norm computation without exploding memory (more details in Section~\ref{sec:ghost-clipping}).

\begin{figure}[t]
  \centering
  \includegraphics[width=\linewidth]{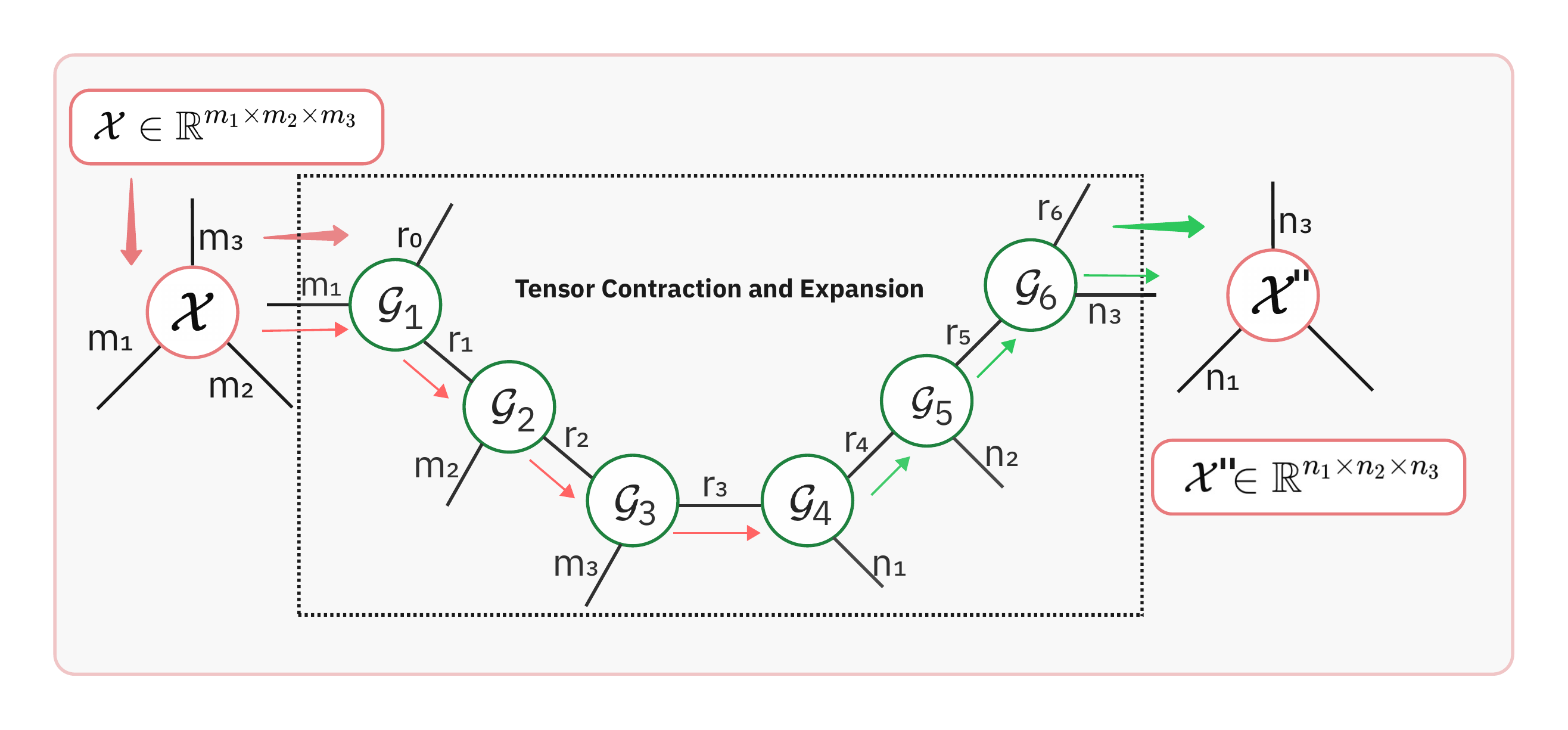}
  \caption{TTLoRA Architecture: Forward Pass Tensor Contraction and Expansion with Input tensor. Red arrow indicates the contraction while green arrow indicates the expansion of the intermediate input activation}
  \label{fig:tensorcontraction}
\end{figure}

\section{Extending Ghost Clipping for TTLoRA-DP}
\label{sec:ghost-clipping}
This section presents our extension of ghost clipping to enable efficient DP-SGD for TTLoRA. The central challenge is computing per-example gradient norms
$\|\nabla_{\theta}\ell_i\|_2$ (where $\ell_i$ is per-example loss)—required for gradient clipping—when the trainable parameters are not a single linear weight matrix but a sequence of Tensor Train (TT) cores.

To address this, we express the TTLoRA forward and backward passes as structured tensor contractions, cache intermediate contraction states, and reuse them to compute \emph{exact} per-example gradient norms without materializing  per-example gradient tensors. This  preserves  DP-SGD privacy guarantees while retaining the computational advantages of ghost clipping.

We proceed in stages. Section~\ref{sec:notation} introduces  notation for TTLoRA-adapted layers and the  forward computation. Section~\ref{sec:inputreshaping} describes reshaping input activation to match the TT factorization. Sections~\ref{sec:contractinginput} and~\ref{sec:expandingoutput} detail the TTLoRA forward pass as a sequence of input contractions and output expansions, during which we cache intermediate states. Section~\ref{sec:persamplebackprop} then shows how these cached states enable efficient per-example gradient norm calculation across TT-cores. Finally, Section~\ref{sec:ghostclipping} integrates the resulting norms into the ghost clipping and gradient update passes, enabling DP-SGD for TTLoRA without excessive memory overhead.

\subsection{Notation}
\label{sec:notation}
Here, we introduce the notation for TTLoRA-adapted layers and the overall forward computation. Let $x\in\mathbb{R}^{B\times S\times d_{\mathrm{in}}}$ be a mini-batch input activation of $B$ sequences of length $S$.
A frozen base linear layer interacting with the input $x$ produces:
\[
y_{\mathrm{base}}(x) \;=\; x\,W^\top,
\qquad
W\in\mathbb{R}^{d_{\mathrm{out}}\times d_{\mathrm{in}}}.
\]
TTLoRA augments this frozen base layer with a trainable TT adapter
$f_{\mathrm{TT}}:\mathbb{R}^{d_{\mathrm{in}}}\rightarrow\mathbb{R}^{d_{\mathrm{out}}}$,
while all base model parameters remain fixed.

\begin{equation}
y(x) \;=\; y_{\mathrm{base}}(x) \;+\; \alpha\, f_{\mathrm{TT}}(x).
\label{eq:overall-y}
\end{equation}

\subsection{Input Reshaping}
\label{sec:inputreshaping}
TTLoRA parameterizes the weight update using TT decomposition, which operates over 3rd-order tensor cores rather than flat vectors. To align the input activations with this structured parameterization, each input vector must first be reshaped into a tensor whose dimensions match the TT factorization of the input cores. This reshaping step is purely structural—it does not change the input values—but enables efficient tensor contractions with TT-cores in the subsequent forward and backward passes.

Let $\mathbf{X} \in \mathbb{R}^{B \times S \times d_{\mathrm{in}}}$ denote a mini-batch of $B$ sequences of length $S$, and let $x_{b,t} \in \mathbb{R}^{d_{\mathrm{in}}}$ denote the input vector at batch index $b$ and token position $t$. Each input vector is reshaped into a $p$-way tensor matching the factorization $d_{\mathrm{in}} = \prod_{k=1}^p m_k$. Formally,
\begin{equation}
\label{eq:xtensor}
    X_{\mathrm{tensor}} \in \mathbb{R}^{B \times S \times m_1 \times \cdots \times m_p},
\end{equation}
where the slice $X_{\mathrm{tensor}}(b,t,:,:,\ldots)$ corresponds to the tensorized form of $x_{b,t}$. This reshaping aligns the input activations with the TT factorization of the adapted weight in TTLoRA.

% An input vector \(\mathbf{X}\in \mathbb{R}^{B\times S\times d_{in}}\) be a batch of \(B\) inputs. Each row of \(\mathbf{X}\) is reshaped into a \(p\)-way tensor matching \([m_1,\dots,m_p]\). Formally,
% \begin{equation}
% \label{eq:xtensor}
%     X_{\text{tensor}} \in \mathbb{R}^{B \times S \times m_1 \times \cdots \times m_p}
% \end{equation}
% so the \(b\)-th row of \(\mathbf{X}\) becomes a \(p\)-dimensional slice in \(\mathbf{X}_{\text{tensor}}\). This reshaping aligns the input with the factors of $d_{in}$ of TT adapted weight during TT factorization used by TTLoRA.

\subsection{Contracting Input Factors}
\label{sec:contractinginput}
We next contract the reshaped input activation with the TT input cores. This step progressively aggregates input dimensions (see Fig.\ref{fig:tensorcontraction}) while introducing TT rank dimensions, and critically, allows us to cache intermediate activation states for further steps. 

We now describe the forward pass for a single token $x_{b,t} \in \mathbb{R}^{d_{\mathrm{in}}}$ from batch index $b$ and sequence position $t$. After reshaping $x_{b,t}$ into a $p$-way tensor with dimensions $(m_1,\dots,m_p)$, we initialize the contraction state:
\[
\mathsf{S}_0(b,t) =x_{b,t} \in \mathbb{R}^{r_0 \times m_p \times \cdots \times m_1},
\qquad r_0 = 1.
\]

We include the trivial rank dimension $r_0=1$ to simplify notation for the subsequent contractions. For $k = 1, \dots, p$, the input contraction between input activation with input TT-cores proceeds sequentially as:
\begin{equation}
\begin{split}
\mathsf{S}_k(b,t) \;=\;
\sum_{a=1}^{r_{k-1}} \sum_{m=1}^{m_k}
\mathsf{S}_{k-1}(b,t)(a,\ldots,m) \times G_k(a,m,\cdot)
\end{split}
\label{eq:forward-input}
\end{equation}
where $G_k \in \mathbb{R}^{r_{k-1} \times m_k \times r_k}$ denotes the $k$-th input TT-core and ${S}_k(b,t)\in \mathbb{R}^{r_k\times m_{k-1}\times \cdots \times m_1}$ denotes the $k$-th intermediate activation state. We cache each pre-core state $\mathsf{S}_{k-1}(b,t) $, which is later reused during per-example backpropagation to compute gradient norms. After contracting all $p$ input cores, the resulting state $\mathsf{S}_p(b,t) \in \mathbb{R}^{r_p}$ provides a compact representation of the input under the TTLoRA adapter.

% We define an initial state $S_{(0)} = X_{\text{tensor}}$ (of equation \ref{eq:xtensor}). For each $k = 1, \ldots, p$, the contracted state $S_{(k)}$ is computed by contracting with TT-core $G_k$:
% \[
% \begin{split}
% S_{(k)}(b, r_k, s, i_{k+1},.., i_p) ={} & \sum_{r_{k-1}=1}^{r_{k-1}} \sum_{i_k=1}^{m_k} S_{(k-1)}(b,r_{k-1},s,i_k,i_{k+1},..,i_p) \\
% & \times G_k(r_{k-1}, i_k, r_k)
% \end{split}
% \]

% for all \(b \in \{1,\dots,B\}\), \(sequence(s) \in \{s_1,\dots,s_S\}\), \(i_k \in \{1,\dots,m_k\}\), and where \(i_{k+1},\dots,i_p\) denote the indices of any remaining (un-contracted) dimensions. We store the \emph{pre-core state} $\mathsf{S}_{k-1}$ before applying $G_k$. In the similar way, we represent the contraction process at token level as well. Let's say, for each token vector $x_{b,t}\in\mathbb{R}^{d_{\mathrm{in}}}$, we reshape and add a rank axis to obtain the initial state:
% \[
% \mathsf{S}_0(b,t)\in\mathbb{R}^{r_0\times m_p\times \cdots \times m_1}
% \quad\text{with } r_0=1,
% \]
% where input modes are ordered in reverse (this matches a performant einsum path). The contraction process of $t^{th}$ token of $b^{th}$ batch will be:

\subsection{Expanding Output Factors}
\label{sec:expandingoutput}
After contracting all input factors, we expand the representation through the output TT-cores to generate the output dimensions (see Fig. \ref{fig:tensorcontraction}). This mirrors the input contraction process and similarly stores intermediate states needed for per-example gradient computation. The contracted state $\mathsf{S}_p(b,t)$ is mapped to the output space using the $q$ output TT-cores $\widetilde{G}_\ell \in \mathbb{R}^{r_{p+\ell-1} \times n_\ell \times r_{p+\ell}}$. We initialize
\[
\mathsf{T}_0(b,t) = \mathsf{S}_p(b,t) \in \mathbb{R}^{r_p}.
\]

For $\ell = 1, \dots, q$, the expansion proceeds sequentially as
\begin{equation}
\mathsf{T}_\ell(b,t)
\;=\;
\sum_{a=1}^{r_{p+\ell-1}}
\mathsf{T}_{\ell-1}(b,t)(a, \ldots)
\, \widetilde{G}_\ell(a, \cdot, \cdot)
\label{eq:tt-output-expansion}
\end{equation}
%where we $\mathsf{T}_\ell(b,t) \in \mathbb{R}^{r_{p+\ell} \times n_1 \times \cdots \times n_\ell}$ and again cache each pre-core state $\mathsf{T}_{\ell-1}(b,t)$. After processing all output cores, the final TT rank satisfies $r_{p+q}=1$, yielding

where $\mathsf{T}_\ell(b,t) \in \mathbb{R}^{r_{p+\ell} \times n_1 \times \cdots \times n_\ell}$, and we again cache each pre-core state $\mathsf{T}_{\ell-1}(b,t)$. After processing all output cores, the final TT rank satisfies $r_{p+q}=1$, yielding

\[
\mathsf{T}_q(b,t) \in \mathbb{R}^{n_1 \times \cdots \times n_q}
\]
Flattening the output modes produces the adapter output $f_{\mathrm{TT}}(x_{b,t}) \in \mathbb{R}^{d_{\mathrm{out}}}$, which is combined with the frozen base layer as
\[
y(x_{b,t}) = x_{b,t} W^\top + \alpha \, f_{\mathrm{TT}}(x_{b,t})
\]

\begin{algorithm}[t]
\caption{Ghost Clipping for TTLoRA}
\label{alg:ghost-clip}
\begin{algorithmic}[1]
\REQUIRE Batch $\{x_b\}_{b=1}^B$, clipping threshold $C$, noise multiplier $\sigma$
\STATE \textbf{Forward Pass:} Compute outputs $\{y_b\}$, cache states $\{S_{(k)}\}$
\STATE \textbf{Loss Computation:} Compute $\mathcal{L}$, get upstream $\{\Delta_b\}$
\STATE \textbf{First Backward (Norm) Pass:}
\FOR{$b = 1$ to $B$}
    \STATE Compute per-example gradients via equations ~\eqref{eq:grad-output-core},~\eqref{eq:grad-input-core}
    \STATE Accumulate $\|g_b\|_2^2$ (discard individual gradients)
\ENDFOR
\STATE \textbf{Compute Clipping Coefficients:} $c_b = \min(1, C/\|g_b\|_2)$
%\STATE \textbf{Reweigh Loss} using clipping coefficient: $\widetilde{l_b}=c_b\cdot l$
\STATE \textbf{Reweight Loss:} $\widetilde{\ell}_b = c_b \cdot \ell_b$
\STATE \textbf{Second Backward (Gradient) Pass:}
\STATE \textbf{Loss backward} assigns clipped gradients to parameters
%\FOR{each TT-core $\theta \in \{G_k, \widetilde{G}_\ell\}$}
%    \STATE $\tilde{g}_\theta = g_\theta + \mathcal{N}(0, \sigma^2 C^2 I)$
%\ENDFOR
\FOR{each TT-core parameter tensor $\theta \in \{G_k, \widetilde{G}_\ell\}$}
    \STATE $\bar{g}_\theta \leftarrow \sum_{b=1}^B c_b\, g_{\theta,b}$
    \STATE $\tilde{g}_\theta \leftarrow \bar{g}_\theta + \mathcal{N}(0, \sigma^2 C^2 I)$
\ENDFOR
%\RETURN Noisy gradients $\{\tilde{g}_\theta\}$

\RETURN Noisy gradients $\{\tilde{g}_\theta\}$
\end{algorithmic}
\end{algorithm}

\subsection{Per-Sample Backpropagation in TT-Cores}
\label{sec:persamplebackprop}
Having expanded the input through the output TT-cores, we now compute per-sample gradients with respect to each TT-core entry. 
Let $\mathcal{L} = \frac{1}{BS} \sum_{b,t} \ell(y_{b,t}, \mathrm{label}_{b,t})$ denote the training loss. 
Autograd provides the upstream gradient for each token:
\[
\Delta(b,t) = \frac{\partial \mathcal{L}}{\partial y_{b,t}} \in \mathbb{R}^{d_{\mathrm{out}}}
\]
Only the TT-adapter path receives the scaled gradient $\alpha \Delta$, which is reshaped to match the output tensor form:
\[
\alpha \Delta(b,t) \;\xrightarrow{\mathrm{reshape}}\; 
\mathsf{C}_q(b,t) \in \mathbb{R}^{r_{p+q} \times n_1 \times \cdots \times n_q}, \quad r_{p+q} = 1
\]

% Let $\mathcal{L}=\frac{1}{BS}\sum_{b,t}\ell(y_{b,t},\mathrm{label}_{b,t})$ be the training loss. Autograd provides the upstream gradient as:
% \[
% \Delta(b,t)\;=\;\frac{\partial \mathcal{L}}{\partial y_{b,t}}\in\mathbb{R}^{d_{\mathrm{out}}}
% \]
% Only the TT path receives the scaled gradient $\alpha\Delta$. We reshape it as:
% \[
% \alpha\Delta(b,t)\;\xrightarrow{\mathrm{reshape}}\;
% \mathsf{C}_q(b,t)\in\mathbb{R}^{r_{p+q}\times n_1\times\cdots\times n_q}
% \quad(r_{p+q}=1),
% \]
which matches $\mathsf{T}_q$. We now propagate per-example gradients right-to-left through the output cores and then through the input cores, reusing cached forward states. Following fast per-example gradient clipping \cite{lee2021scaling}, we compute per-token gradients in factored form as (upstream)$\times$(activation), and then apply this principle to TT-core contractions using cached states. %We adopt the per-sample gradient formulation proposed in Fast per-example gradient clipping \cite{lee2021scaling}, where the gradient of the loss with respect to the weight \( W_{i,j} \) is computed as:
\[
\left( \frac{\partial L}{\partial W_{i,j}} \right) = \frac{\partial L}{\partial Y_{i}} \cdot X_{j}
\]
% For compactness, we define prefix/suffix products as:
% \[
% Z^{\mathrm{in}}_{<k} \;=\; \prod_{u=1}^{k-1} m_u,\quad
% Z^{\mathrm{out}}_{<\ell} \;=\; \prod_{v=1}^{\ell-1} n_v,\quad
% Z^{\mathrm{out}}_{>\ell} \;=\; \prod_{v=\ell+1}^{q} n_v,
% \]
% where $Z^{\mathrm{in}}_{<k}$ is the product of all input modes $m_u$ where $u$ is less than the current index $k$, $Z^{\mathrm{out}}_{<\ell}$ is the product of all output modes $n_v$ where $v$ is less than the current index $l$ and $Z^{\mathrm{out}}_{>\ell}$ is the product of all output modes $n_v$ where $v$ is greater than the current index $l$. And the empty products equal to $1$.

\subsubsection{\textbf{Gradients w.r.t.\ output cores $\widetilde G_\ell$.}}
Let $\mathsf{T}_{\ell-1}(b,t)$ be the pre-core state stored in the forward pass before contracting it with $\widetilde G_\ell$. The per-sample, per-token gradient is:
\begin{equation}
    \frac{\partial \ell_{b,t}}{\partial \widetilde{G}_\ell(a, n, b')} = \sum_{\text{context}} C_\ell(b,t)[b', \ldots, n, \ldots] \cdot T_{\ell-1}(b,t)[a, \ldots]
    \label{eq:grad-output-core}
\end{equation}
which has shape $r_{p+\ell-1}\times n_\ell\times r_{p+\ell}$. Here, \emph{context} denotes the product of all remaining indices in $n_1,\ldots,n_q$ except $n_\ell$ (and any rank indices implied by the current state).
Backpropagating to the input of core $\ell$, which we store as the upstream gradient for the next core, yields:
\begin{equation}
    C_{\ell-1}(b,t)[a, \ldots] = \sum_{n, b'} C_\ell(b,t)[b', \ldots, n, \ldots] \cdot \widetilde{G}_\ell(a, n, b')
    \label{eq:upstream-output-core}
\end{equation}

\subsubsection{\textbf{Gradients w.r.t.\ input cores $G_k$.}}
After processing all output cores, we obtain $\mathsf{C}_0(b,t)\in\mathbb{R}^{r_p}$, the upstream for the input cores. For $k=p,p-1,\dots,1$, let the forward pre-core state be
$\mathsf{S}_{k-1}(b,t)$ and $\mathsf{U}_k(b,t)$ be the current upstream (gradient w.r.t.\ $\mathsf{S}_k$). Then per-sample, per-token gradient is,
\begin{equation}
    \frac{\partial \ell_{b,t}}{\partial G_k(a, m, b')} = \sum_{z} U_k(b,t)[b', z] \cdot S_{k-1}(b,t)[a, z, m]
    \label{eq:grad-input-core}
\end{equation}

and the upstream for the next core, which we store for the upcoming core, is:
\begin{equation}
\begin{split}
\mathsf{U}_{k-1}(b,t)[a,z,m]
\;=\;
\sum_{b'=1}^{r_k}
\mathsf{U}_k(b,t)[b',z] \\
\times G_k(a,m,b')\;
\end{split}
\label{eq:upstream-input-core}
\end{equation}
For $k>1$, $\mathsf{U}_{k-1}$ matches the shape of $\mathsf{S}_{k-1}$ and becomes the next upstream.

\subsection{Ghost Clipping on Per-Sample Gradients}
\label{sec:ghostclipping}
To efficiently enforce per-example $\ell_2$ clipping while keeping memory usage manageable, we implement a two-pass procedure from Ghost Clipping. Algorithm~\ref{alg:ghost-clip} summarizes this procedure, showing how forward caching, per-example norm computation, and a second backward pass with gradient scaling and noise addition are orchestrated.

With per-sample gradients for all TT-core parameters computed in above sections, we now enforce $\ell_2$-norm bounds for DP. This ensures that each example contributes at most $C$ to the gradient, which is critical for differential privacy and for preventing outlier gradients from destabilizing training. Let $\theta$ be any trainable parameter (a TT-core entry). For example $(b,t)$ the per-sample gradient is $\nabla_{\!\theta}\ell_{b,t}$ computed by~\eqref{eq:grad-output-core} and~\eqref{eq:grad-input-core}. For each parameter, we accumulate each parameter's per example norm contribution across tokens and cores as:
\[
g^{(b)}_\theta
\;=\;
\Big\|
\big(\nabla_{\!\theta}\ell_{b,1},\dots,\nabla_{\!\theta}\ell_{b,S}\big)
\Big\|_2
\quad \text{(aggregated over tokens)}
\]
and combine all parameters to form a per-example global norm:
\[
\|g_b\|_2
\;=\;
\Bigg(\sum_{\theta}\big(g^{(b)}_\theta\big)^2\Bigg)^{1/2}
\]
Clipping coefficients are then applied in a second backward pass, scaling each example’s contribution so that the overall per-example gradient is $\ell_2$-bounded by $C$ as (small value $\tau \sim 10^{-6}$ is added to avoid division by 0):
\[
c_b \;=\;\min\!\Big\{1,\;\tfrac{C}{\|g_b\|_2+\tau}\Big\}
\]

% \textbf{Two-Pass Algorithm:} 
% \begin{enumerate}
%     \item \textbf{Norm Pass:} Compute per-example gradient norms using cached states and equations~\eqref{eq:grad-output-core},~\eqref{eq:grad-input-core}, accumulating $\|g_b\|_2$ without storing full gradients.
%     \item \textbf{Gradient Pass:} Recompute gradients, scale by $c_b$, aggregate across examples, and add calibrated Gaussian noise.
% \end{enumerate}

\textbf{Memory Complexity:} Materializing per-example gradients requires $\mathcal{O}(B \cdot |\theta|)$ memory where $|\theta|$ is the parameter count. Our ghost clipping implementation avoids storing per-example gradients and instead caches only TT contraction states requiring  only $\mathcal{O}(BS \cdot r^2 \cdot (p+q))$ memory for cached states plus $\mathcal{O}(B)$ for per-example norms.
In particular, memory does not scale with the size of the implicit matrix $\Delta W \in \mathbb{R}^{d_{\mathrm{out}}\times d_{\mathrm{in}}}$.

\section{Analysis of Privacy–Utility Trade-offs in TT-LoRA}
\label{sec:analysis}
In this section, we evaluate TTLoRA’s privacy–utility profile relative to  LoRA and full fine-tuning (FFT) under both  DP-SGD and non-private training. We first describe the model, datasets, PEFT configurations, and evaluation setup (Section~ \ref{subsec:modelanddataset}), followed by the the details about PreCurious attack protocol (Section \ref{subsec:mia-precurious}). Furthermore, we systematically study TTLoRA's privacy--utility properties along three complementary dimensions:
\begin{enumerate}
    \item \textbf{Utility under Differential Privacy:} We evaluate how DP-SGD affects language-modeling performance across privacy budgets, comparing privately trained TTLoRA, LoRA, and FFT (Sec.~\ref{sec:privacy-utility}, Table~\ref{tab:dp-ppl}, Fig.~\ref{fig:dp-perplexity},~\ref{fig:2boxplot}).

    \item \textbf{Membership Inference Vulnerability under DP:} We analyze how DP budgets influence susceptibility to PreCurious membership inference attack, highlighting TTLoRA's robustness relative to LoRA and FFT 
    (Sec.~\ref{sec:privacy-utility-attack}, Table~\ref{tab:dp-mia-avg-comparison}, Fig.~\ref{fig:3dplot},~\ref{fig:privacy-gain}).

    \item \textbf{Inherent Privacy Resilience without DP:} We examine TTLoRA's intrinsic privacy properties in a non-private setting, showing that even without DP, TTLoRA substantially reduces membership leakage while maintaining competitive utility 
    (Sec.~\ref{sec:nonprivatebaseline}, Table~\ref{tab:nonprivate}, Fig.~\ref{fig:loss-distribution}, ~\ref{fig:privacy-utility-nonprivate}).
\end{enumerate}

Our experiments reveal three key insights: (1) under identical DP constraints, TTLoRA matches or improves utility relative to LoRA while using an order-of-magnitude fewer trainable parameters; (2) TTLoRA exhibits lower membership leakage preserving utility under DP training, remaining nearly flat as $\varepsilon$ increases, whereas LoRA’s leakage grows substantially; (3) under non-private training, TTLoRA exhibits substantially lower membership leakage in exchange of less utility cost; these findings suggest that TTLoRA’s tensor-train structure can materially reduce the membership signal exploited by loss-based attacks.

\subsection{Model and Dataset}
\label{subsec:modelanddataset}
Following prior works \cite{liu2024precurious, malekmohammadi2024low}, we fine-tune GPT-2 (124M) on two benchmark datasets containing confidential or sensitive content:
\begin{itemize}
    \item \textbf{Enron Email Corpus} \cite{klimt2004enron}: Corporate email communications released during the Enron investigation, containing sensitive business discussions.
    \item \textbf{Penn Treebank (PTB)} \cite{marcus1993building}: Wall Street Journal articles covering business and financial news.
\end{itemize}
We evaluate utility using validation perplexity (PPL; lower is better) and privacy leakage using attack AUC-ROC (lower is better) along with different recall rates at variable False Positive Rates (FPR) (lower is better). \\
\textbf{Fine-Tuning Methods:} We compare three approaches:
\begin{itemize}
    \item \textbf{Full Fine-Tuning (FFT):} All 124M parameters trainable; Learning rate $1e^{-4}$
    \item \textbf{LoRA:} Adapted on attention weights (c\_attn) and projection weights (c\_proj) of all 12 layers with ranks $r \in \{2, 4, 6, 8, 10, 12, 14, 16\}$, yielding 110.6K--884.7K trainable parameters; Learning rate $5e^{-4}$
    \item \textbf{TTLoRA:} Tensor train adapters on the same weights as LoRA adaptation with matching ranks, yielding 7.6K--144.4K trainable parameters. The TT-factorization uses $[64, 4, 3]$ factors for 768 and $[64, 4, 3, 3]$ factors for 2304 and factorize c\_attn with dimension 768$\times$2304 and c\_proj with dimension 768$\times$768; Learning rate $5e^{-3}$
\end{itemize}

\paragraph{\textbf{Privacy Configurations:}} We train under four privacy budgets $\varepsilon \in \{0.5, 1.0, 3.0, 5.0\}$ and report a non-private baseline (DP disabled). For DP runs, we use $\delta = N^{-1.1}$ where $N$ is the number of training samples, consistent with prior practice. For DP training, we use private-transformers\footnote{https://github.com/lxuechen/private-transformers} \cite{li2021large} with our TTLoRA ghost clipping extension with effective batch size 32, and tune the noise multiplier to achieve the target $\varepsilon$ over the training steps using Renyi DP accounting method \cite{mironov2017renyi}.

\paragraph{\textbf{Evaluation Metrics:}} We measure privacy and utility metrics: 
\begin{itemize}
    \item \textbf{Utility:} Validation perplexity (PPL, lower is better),
    \item \textbf{Privacy Leakage:} We evaluate membership inference attack success using the calibrated loss-threshold method PreCurious (more details in section \ref{subsec:mia-precurious}) \cite{liu2024precurious}. We measure: (1) \emph{Attack AUC}: area under the ROC curve (50\% = random guess), (2) \emph{TPR@FPR=1\%}: True Positive Rate at False Positive Rate of 0.01, (3) \emph{TPR@FPR=0.1\%}: True Positive Rate at False Positive Rate of 0.001, (4) \emph{TPR@FPR=0.01\%}: True Positive Rate at False Positive Rate of 0.0001.
\end{itemize}

\paragraph{\textbf{Ranks Configuration:}}
All experiments for both LoRA and TTLoRA are conducted using even ranks $r \in \{2,4,\ldots,16\}$, and we report rank-averaged results in Tables~\ref{tab:dp-ppl}, \ref{tab:dp-mia-avg-comparison}, and~\ref{tab:nonprivate}. 
This range corresponds to the low to moderate rank regime commonly explored in prior PEFT studies~\cite{hu2022lora,ma2024efficient,biderman2024lora, kunwar2025tt}, where adapters provide substantial parameter efficiency without approaching the behavior of full fine-tuning. Larger ranks increasingly saturate utility while substantially increasing the number of trainable parameters and exhibit increasing vulnerability towards MIA leakage (see Table 1), whereas very small ranks often suffer from unstable training and degraded performance.

To ensure that our conclusions are not sensitive to a particular rank choice, we average results across this range. This reporting strategy mitigates variance due to rank-specific effects and enables a robust comparison between LoRA and TTLoRA. Full rank-wise results for Section \ref{sec:privacy-utility}, 
\ref{sec:privacy-utility-attack}, \ref{sec:nonprivatebaseline} are provided in the appendix due to space constraints. 
% We selected the rank range $r\in[2,16]$ to systematically characterize the Pareto frontier between structural regularization and model expressivity, spanning the spectrum from extreme constraint ($r=2$) to a saturation point ($r=16$) where additional complexity yields diminishing utility returns while increasing empirical leakage (Appendix E, Table 5). We report results averaged across these ranks to demonstrate the methodological robustness of TTLoRA-DP, proving that its privacy-utility advantages are intrinsic to the TT structure's inductive bias against memorization rather than artifacts of brittle hyperparameter tuning. This approach provides a rigorous validation of the framework's stability, confirming that the structural bottleneck remains effective across a broad operational regime without relying on precise, budget-consuming configuration searches.

\begin{table}[htbp]

  \centering
  \caption{\small Comparison of attack metrics (AUC, FPR@1\%, FPR@0.01\%) and utility (PPL) for LoRA, TTLoRA, and FFT on the Enron dataset under non-private (no DP) setting. Lower ↓ is better. Bold orange (LoRA) and blue (TTLoRA) indicate best individual performance; bold black indicates best average performance; gray highlighted rows correspond to equivalent parameter count configurations.}
\label{tab:nonprivate-enron-only}
  \setlength{\tabcolsep}{3.0pt}
  \begin{tabular}{p{0.90cm} c c c c c c}
    \toprule
    Method & Rank & Params 
    & PPL 
    & AUC 
    & FPR@1\% 
    & FPR@0.01\% \\
    \midrule

    \rowcolor{gray!15}
    & 2  & 110.6k & 17.74 & \textbf{\textcolor{myorange}{92.23\%}} & \textbf{\textcolor{myorange}{25.34\%}} & \textbf{\textcolor{myorange}{0.45\%}} \\
    & 4  & 221.2k & 17.59 & 94.71\% & 35.23\% & 0.90\% \\
    & 6  & 331.8k & 17.41 & 95.41\% & 36.28\% & 0.75\% \\
    & 8  & 442.4k & 17.62 & 93.89\% & 37.33\% & 1.95\% \\
    \textbf{LoRA}
    & 10 & 553.0k & 17.36 & 96.04\% & 27.29\% & 1.65\% \\
    & 12 & 663.6k & 17.38 & 95.70\% & 37.78\% & 2.40\% \\
    & 14 & 774.1k & \textbf{\textcolor{myorange}{17.29}} & 95.95\% & 28.49\% & 1.35\% \\
    & 16 & 884.7k & 17.32 & 95.35\% & 40.03\% & 1.50\% \\
    \cmidrule(lr){2-7}
    \textbf{Avg.} & -- & 497.7k & \textbf{17.46} & 94.91\% & 33.47\% & 1.37\% \\

    \midrule

    & 2  & 7.6k  & 18.35 & \textbf{\textcolor{myblue}{71.49\%}} & \textbf{\textcolor{myblue}{4.20\%}} & 0.45\% \\
    & 4  & 18.2k & 18.23 & 84.35\% & 5.25\% & 1.65\% \\
    & 6  & 31.8k & 18.11 & 88.72\% & 13.94\% & 1.80\% \\
    & 8  & 48.4k & 17.97 & 92.99\% & 21.59\% & 0.75\% \\
    \textbf{TTLoRA}
    & 10 & 67.9k & 18.04 & 92.30\% & 18.74\% & 1.95\% \\
    \rowcolor{gray!15}
    & 12 & 90.4k & 17.90 & 93.85\% & 19.94\% & \textbf{\textcolor{myblue}{0.30\%}} \\
    \rowcolor{gray!10}
    & 14 & 115.9k & 17.97 & 91.65\% & 22.04\% & 1.65\% \\
    & 16 & 144.4k & \textbf{\textcolor{myblue}{17.83}} & 94.10\% & 23.84\% & \textbf{\textcolor{myblue}{0.30\%}} \\
    \cmidrule(lr){2-7}
    \textbf{Avg.} & -- & \textbf{65.6k} & 18.05 & \textbf{88.68\%} & \textbf{16.19\%} & \textbf{1.11\%} \\

    \midrule
    \textbf{FFT} & -- & 124.04M & 16.45 & 96.63\% & 26.24\% & 4.35\% \\
    \bottomrule
  \end{tabular}
\end{table}

\subsection{PreCurious Membership Inference Attack}
\label{subsec:mia-precurious}
Following the \textsc{PreCurious} threat model~\cite{liu2024precurious}, we consider a prominent black-box adversary who (i) publishes a crafted pre-trained checkpoint (ii) later queries the victim’s released fine-tuned model. The attack can be viewed in three stages: 

\paragraph{\textbf{Crafting:}} The adversary \textbf{warms-up} a benign pre-trained model $\theta_{\text{benign}}$ (fully fine-tune for some epochs) on an auxiliary corpus ($D_{\mathrm{aux}}$: drawn from the same distribution as the victim’s private fine-tuning data but disjoint from it) to obtain an adversarial initialization and later uses as a reference model $\theta_{\text{ref}}$.

\paragraph{\textbf{Fine-tuning:}} The victim downloads the released checkpoint $\theta_{\text{ref}}$, attaches PEFT modules, fine-tunes on private data $D_{train}$ and releases $\theta_{\text{peft}}$ model, 
\paragraph{\textbf{Inferring:}} The adversary performs membership inference using only loss queries. We split the original training corpus into three equal partitions $(D_{\text{aux}}, D_{\text{train}}, D_{\text{non}})$ to mimic the real-world scenario, where $D_{\text{non}}$ serves as a non-member calibration set. At inference time, we compute a calibrated membership score using the crafted reference model $\theta_{\text{ref}}$ trained on $D_{\text{aux}}$ as:

\begin{equation}
s_{\text{ref}}(x) = \mathcal{L}(x;\theta_{\text{peft}}) - \mathcal{L}(x;\theta_{\text{ref}}),
\label{eq:precurious-ref-score}
\end{equation}
where $\mathcal{L}$ is the per-sample loss. Given a target false-positive rate $\alpha$, we set the threshold $\tau_\alpha$ as the largest value that keeps the FPR on $D_{\text{non}}$ below $\alpha$ such that $\Pr_{x\sim D_{\mathrm{non}}}\!\left[s_{\mathrm{ref}}(x)\le \tau_\alpha\right]=\alpha$ and $\mathrm{TPR}(\alpha)=\Pr_{x\sim D_{\mathrm{train}}}\!\left[s_{\mathrm{ref}}(x)\le \tau_\alpha\right]$. We summarize attack effectiveness using ROC-AUC by treating members as label 1 and non-members as label 0, and using scores $-s_{\text{ref}}(x)$ so that lower calibrated loss corresponds to higher membership likelihood.

\subsection{Differentially Private (DP) Training: Privacy–Utility Trade-offs}
\label{sec:privacy-utility}
We evaluate the utility of DP fine-tuning under DP-SGD on Enron and PTB datasets.
Unlike our attack evaluation protocol (which splits the corpus into $D_{\mathrm{aux}},D_{\mathrm{train}},D_{\mathrm{non}}$), here we fine-tune on the \emph{full} training set to isolate the privacy--utility impact of DP-SGD.
We adapt a pretrained GPT-2 checkpoint using either LoRA or TTLoRA and train for 15 epochs under privacy budgets $\varepsilon\in\{0.5,1.0,3.0,5.0\}$, selecting the best validation checkpoint (by validation PPL) for each $\varepsilon$ for reporting.
For LoRA and TTLoRA, we extract results for even ranks $r\in\{2,4,\dots,16\}$; Table~\ref{tab:dp-ppl} reports results averaged over ranks, while Table~\ref{tab:comprehensive_ppl_even} in appendix provides the full rank-wise breakdown.

The non-private column in Table~\ref{tab:dp-ppl} is obtained by running the DP training pipeline with DP disabled, and is not directly comparable to the non-private column of Table~\ref{tab:nonprivate} which is measured under the attack pipeline and uses a different data split and warm-start procedure following the PreCurious protocol.

\begin{table}[h]
\centering
\caption{Perplexity (lower is better) under DP averaged over performance at even ranks spanning 2 to 16. TTLoRA achieves better utility than LoRA at all privacy budgets.}
\label{tab:dp-ppl}
\small
\begin{tabular}{lcccccc}
\toprule
\textbf{Method}& \textbf{Params} & \multicolumn{4}{c}{\textbf{Privacy Budget ($\varepsilon$)}} & \textbf{Non-Priv.} \\
\cmidrule(lr){3-6}
& & 0.5 & 1.0 & 3.0 & 5.0 & \\
\midrule
\multicolumn{7}{c}{\textit{Enron Dataset}} \\
\midrule
FFT & 124.0M & 28.81 & 27.05 & 25.38 & 24.92 & \textbf{14.31} \\
LoRA (avg.) & 497.7K & 29.29 & 27.62 & 26.64 & 26.44 & 18.79 \\
TTLoRA (avg.) & 65.6K & \textbf{27.52} & \textbf{26.55} & \textbf{25.37} & \textbf{24.99} & 20.14 \\
\midrule
\multicolumn{7}{c}{\textit{PTB Dataset}} \\
\midrule
FFT & 124.0M & \textbf{56.18} & \textbf{53.38} & \textbf{49.23} & \textbf{47.79} & \textbf{19.75} \\
LoRA (Avg.) & 497.7K & 60.41 & 58.59 & 54.75 & 53.81 & 28.86 \\
TTLoRA (Avg.) & 65.6K & 58.47 & 55.39 & 50.61 & 48.71 & 31.45 \\
\bottomrule
\end{tabular}
\end{table}

\begin{figure}[t]
    \centering
    \includegraphics[width=\linewidth]{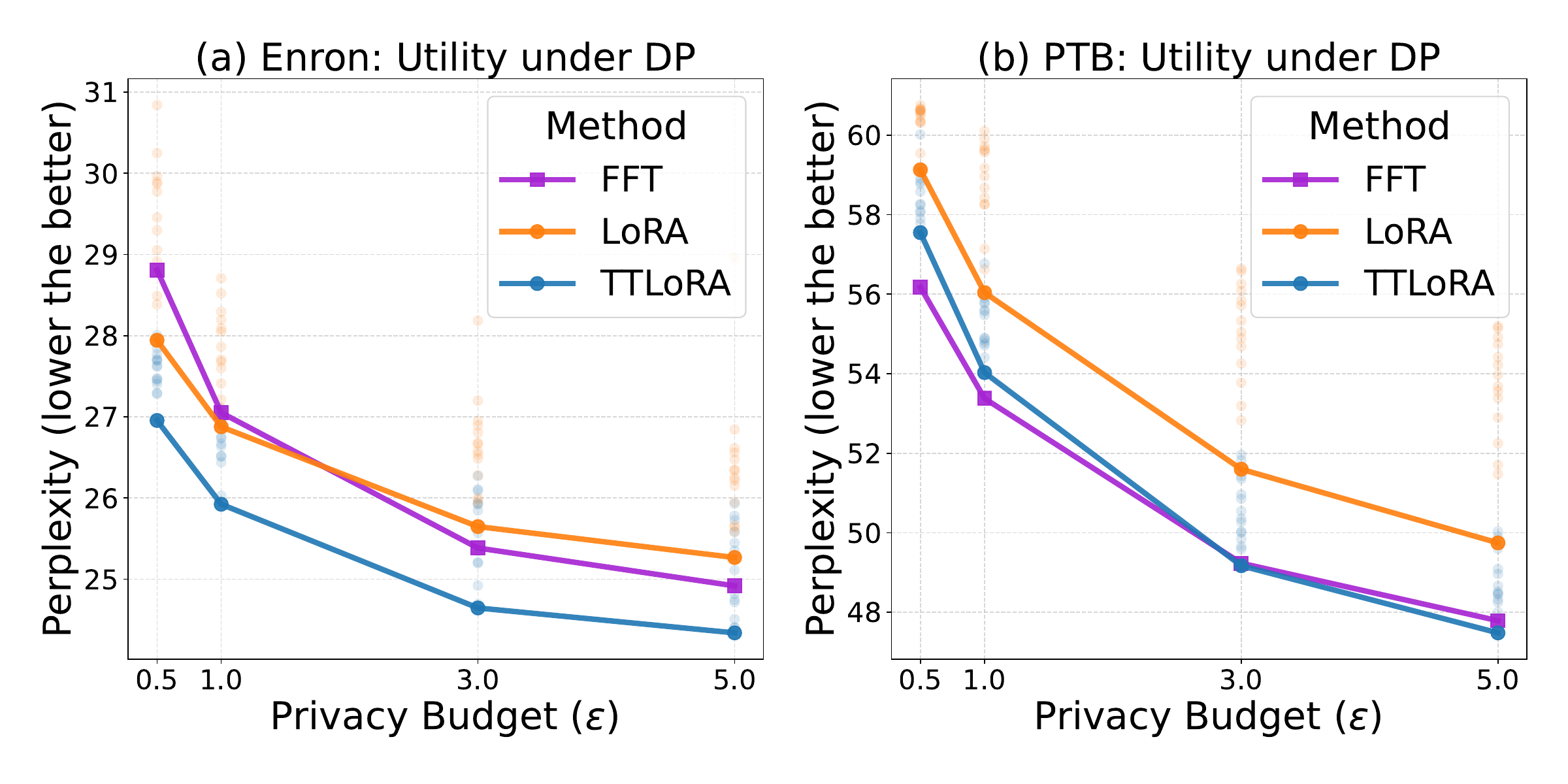}
    \caption{\textbf{Pareto Frontier: Best utility extracted among multiple configurations under differential privacy.} TTLoRA achieves better perplexity than LoRA at all privacy budgets, with the advantage strongest at stricter (lower $\varepsilon$) settings.}
    \label{fig:dp-perplexity}
\end{figure}

\begin{figure}[t]
  \centering
  \includegraphics[width=\linewidth]{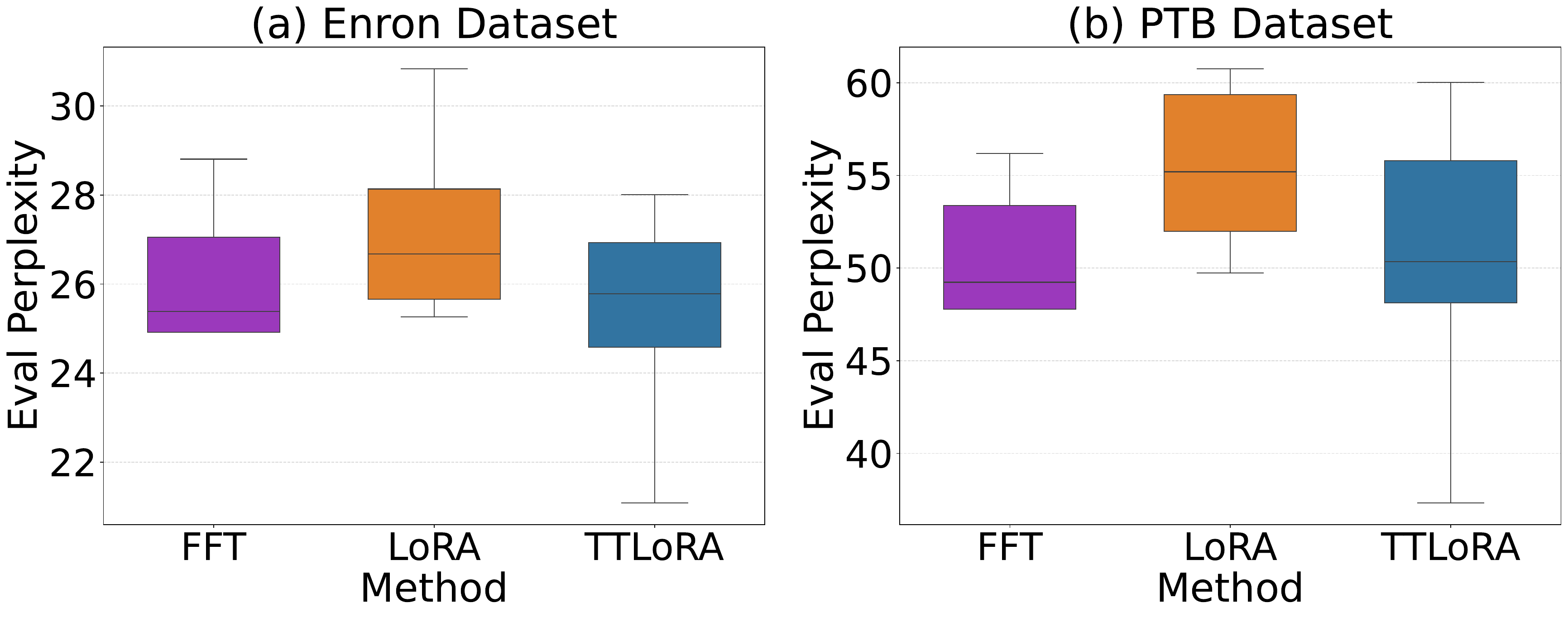}
  \caption{Overall utility comparison over different rank configurations for LoRA and TTLoRA along with FFT under DP training with Enron and PTB datasets}
  \label{fig:2boxplot}
\end{figure}

We observe that \textbf{TTLoRA improves DP utility despite fewer parameters.} Across all privacy budgets, TTLoRA achieves lower (better) perplexity than LoRA on both datasets while using substantially fewer trainable parameters (65.6K vs.\ 497.7K on average).
On Enron, TTLoRA improves over LoRA at every $\varepsilon$: e.g., at $\varepsilon{=}0.5$, TTLoRA attains 27.52 PPL vs.\ 29.29 for LoRA, and at $\varepsilon{=}3.0$ it attains 25.37 vs.\ 26.64.
On PTB, the gap is even larger: at $\varepsilon{=}5.0$, TTLoRA achieves 48.71 vs.\ 53.81 for LoRA.
Figure~\ref{fig:dp-perplexity} (which plots the best utility among 16 ranks configurations) visualizes the same trend as a function of $\varepsilon$: utility improves monotonically as the privacy budget increases (less noise), and TTLoRA consistently traces a curve closer to or even better than FFT than LoRA.

\textbf{Relationship to full fine-tuning (FFT).} FFT remains the strongest non-private baseline in perplexity, but under DP-SGD its advantage narrows substantially.
On Enron, TTLoRA matches or slightly exceeds FFT at moderate privacy (25.37 vs.\ 25.38 at $\varepsilon{=}3.0$), and outperforms FFT at strict privacy (27.52 vs.\ 28.81 at $\varepsilon{=}0.5$), despite using orders of magnitude fewer trainable parameters.
On PTB, FFT yields the best perplexity across budgets, but TTLoRA closes most of the gap relative to LoRA (e.g., at $\varepsilon{=}5.0$, TTLoRA is within $\approx$0.92 PPL of FFT, whereas LoRA is $\approx$6.02 PPL worse).
The box-plot summary in Figure~\ref{fig:2boxplot} further corroborates this: TTLoRA exhibits consistently lower median perplexity and reduced dispersion compared to LoRA under DP.

\subsection{DP Training under Attack: MIA at Varying Privacy Budgets}
\label{sec:privacy-utility-attack}
We evaluate privacy leakage under differentially private (DP) fine-tuning using the \emph{PreCurious} membership inference attack~\cite{liu2024precurious}. For each dataset, we partition the original training split into three equal parts: an auxiliary subset $D_{\text{aux}}$ for warm-up, a member subset $D_{\text{train}}$ used for DP fine-tuning, and a disjoint non-member subset $D_{\text{non}}$ used only for attack calibration. We warm up a GPT-2 base model for 2 epochs on $D_{\text{aux}}$ to obtain an adversarially-initialized checkpoint, and set the resulting model as the reference $\theta_{\text{ref}}$. We then adapt the model with either LoRA or TTLoRA and fine-tune the PEFT parameters on $D_{\text{train}}$ for 15 epochs with DP-SGD, targeting privacy budgets $\varepsilon\in\{0.5,1.0,3.0,5.0\}$ (fixed $\delta$ and clipping configuration across methods).

Following PreCurious, we compute calibrated membership scores based on reference vs. adapted losses using equation~\ref{eq:precurious-ref-score} and summarize attack strength using ROC-AUC and fixed-FPR operating points (here reported as FPR@10\%).
For each $(\text{method}, \text{rank}, \varepsilon)$ configuration, we report the \emph{best epoch} (based on validation perplexity and extract the result for all $\varepsilon$ for the same epoch) and aggregate results over even ranks $r\in\{2,4,\dots,16\}$ as presented in Table ~\ref{tab:dp-mia-avg-comparison} while Table \ref{tab:mia-dp-enron-only} and \ref{tab:mia-dp-ptb-only} in appendix present the detailed rank-wise performance for Enron and PTB datasets. 

\begin{table*}[h]
\centering
\caption{Comparison of attack metrics (AUC$\downarrow$ and TPR@FPR=10\%$\downarrow$) and utility (PPL$\downarrow$) for LoRA, TTLoRA (averaged over ranks 2--16), and FFT across Enron and PTB datasets under DP-SGD. Lower is better for AUC, TPR@FPR=10\%, and PPL.}
\label{tab:dp-mia-avg-comparison}
\small
\begin{tabular}{l ccc ccc ccc ccc}
\toprule
& \multicolumn{3}{c}{$\varepsilon=0.5$} & \multicolumn{3}{c}{$\varepsilon=1.0$} & \multicolumn{3}{c}{$\varepsilon=3.0$} & \multicolumn{3}{c}{$\varepsilon=5.0$} \\
\cmidrule(lr){2-4} \cmidrule(lr){5-7} \cmidrule(lr){8-10} \cmidrule(lr){11-13}
\textbf{Method} & AUC & PPL & FPR@10\% & AUC & PPL & FPR@10\% & AUC & PPL & FPR@10\% & AUC & PPL & FPR@10\% \\
\midrule
\multicolumn{13}{c}{\textbf{Enron Dataset}} \\
\midrule
FFT & 54.00\% & 21.48 & 11.24\% & 55.06\% & 21.25 & 11.84\% & 57.66\% & 20.99 & 15.29\% & 59.04\% & 20.90 & 15.74\% \\
LoRA (avg.) & 52.49\% & \textbf{20.72} & 11.14\% & 53.72\% & \textbf{20.72} & 12.11\% & 56.99\% & \textbf{20.71} & 14.91\% & 58.43\% & \textbf{20.71} & 15.89\% \\
TTLoRA (avg.) & \textbf{51.36\%} & \textbf{20.72} & \textbf{10.66\%} & \textbf{51.48\%} & \textbf{20.72} & \textbf{10.87\%} & \textbf{51.76\%} & 20.72 & \textbf{11.19\%} & \textbf{52.04\%} & 20.72 & \textbf{11.23\%} \\
\midrule
\multicolumn{13}{c}{\textbf{PTB Dataset}} \\
\midrule
FFT & 50.09\% & 32.52 & 10.63\% & 51.74\% & 32.36 & 11.49\% & 56.32\% & 32.05 & 15.23\% & 59.28\% & 31.90 & 19.25\% \\
LoRA (avg.) & 52.62\% & \textbf{31.51} & 10.88\% & 53.57\% & \textbf{31.51} & 12.18\% & 56.55\% & \textbf{31.51} & 15.27\% & 58.58\% & \textbf{31.51} & 18.14\% \\
TTLoRA (avg.) & \textbf{51.62\%} & 31.52 & \textbf{9.41\%} & \textbf{51.65\%} & 31.52 & \textbf{9.34\%} & \textbf{51.82\%} & \textbf{31.51} & \textbf{9.45\%} & \textbf{52.19\%} & \textbf{31.51} & \textbf{9.84\%} \\
\bottomrule
\end{tabular}
\end{table*}

\begin{figure}[h]
    \centering
    \includegraphics[width=\linewidth]{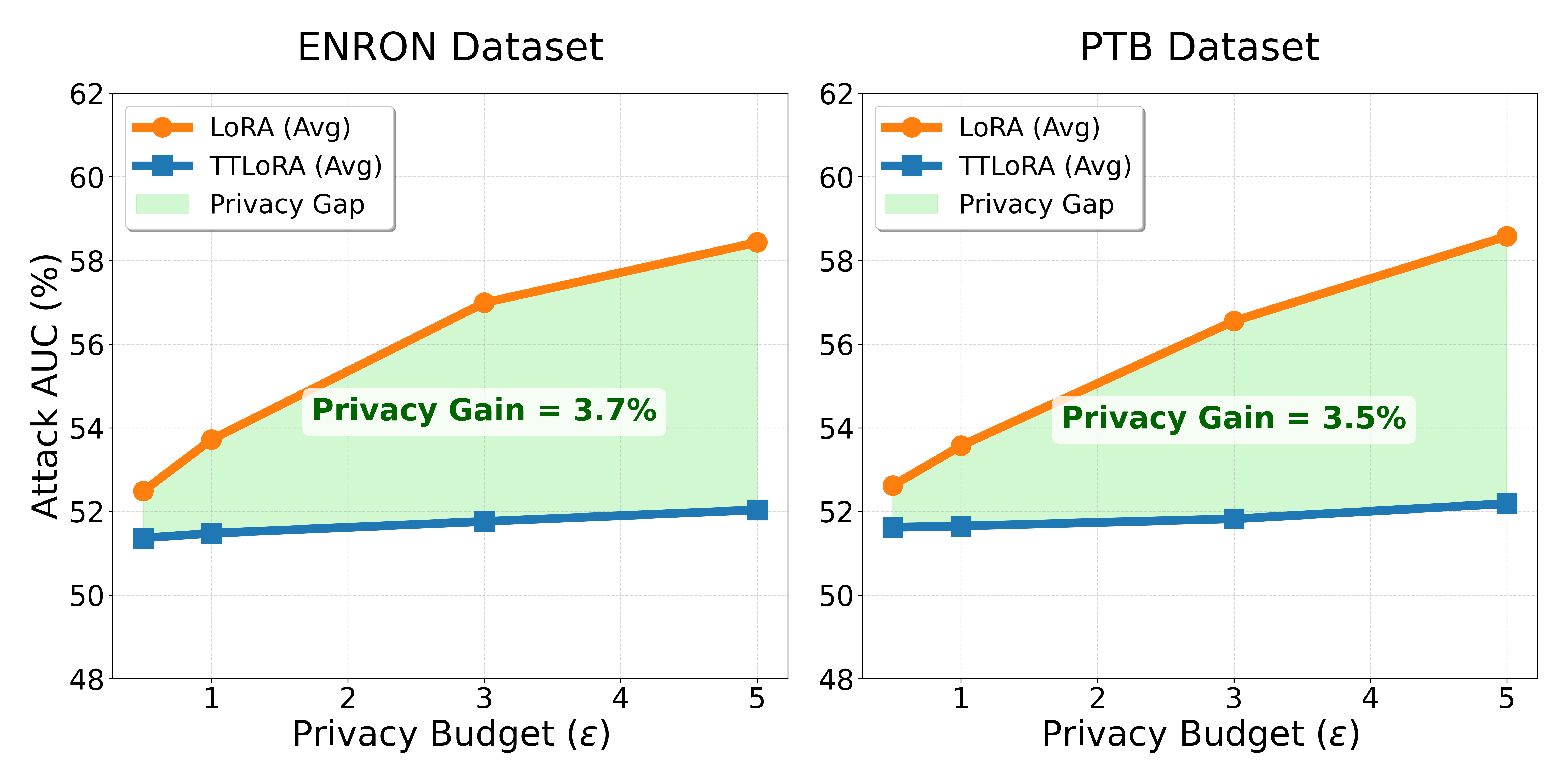}
    \caption{\textbf{MIA AUC under PreCurious (rank-averaged).} TTLoRA remains near the random-guess baseline (50\%) across all $\varepsilon$, while LoRA becomes increasingly vulnerable as $\varepsilon$ increases. Shaded region denotes the AUC gap (mean privacy gain: 3.7 points on Enron, 3.5 points on PTB).}
    \label{fig:privacy-gain}
\end{figure}

\textbf{TTLoRA is consistently harder to infer under DP, and the advantage widens as $\varepsilon$ increases.}
Across both Enron and PTB, TTLoRA yields uniformly lower attack AUC than LoRA at the same privacy budget. On Enron, LoRA’s rank-averaged AUC increases sharply as the privacy constraint relaxes (52.49\% $\rightarrow$ 58.43\% from $\varepsilon=0.5$ to 5), while TTLoRA remains close to the 50\% random-guess baseline and changes only marginally (51.36\% $\rightarrow$ 52.04\%). This produces a growing AUC gap from 1.13 points at $\varepsilon=0.5$ to 6.39 points at $\varepsilon=5$. A similar trend holds on PTB, where LoRA rises from 52.62\% to 58.58\%, whereas TTLoRA stays near 51.62--52.19\%. Figure~\ref{fig:privacy-gain} visualizes this effect and reports the average AUC reduction achieved by TTLoRA across $\varepsilon$ (3.7 points on Enron and 3.5 points on PTB).

\textbf{Fixed-FPR operating points show increasing separability for LoRA but near-constant leakage for TTLoRA.}
The same widening pattern appears under $\mathrm{TPR}@\mathrm{FPR}=10\%$. On Enron, LoRA’s $\mathrm{TPR}@10\%$ increases from 11.14\% to 15.89\% as $\varepsilon$ increases, whereas TTLoRA remains nearly flat (10.66\% to 11.23\%). On PTB, LoRA reaches 18.14\% at $\varepsilon=5$, while TTLoRA stays below 10\% across all budgets (9.34--9.84\%). These results indicate that relaxing the privacy budget (adding less noise) significantly increases membership distinguishability for LoRA, but has limited effect for TTLoRA.

\textbf{Utility is essentially matched in the attack-evaluation pipeline, so privacy gains are not driven by degraded accuracy.}
In the PreCurious setting, LoRA and TTLoRA achieve nearly identical rank-averaged perplexity and exhibit minimal sensitivity to $\varepsilon$ (e.g., Enron $\approx$20.71--20.72; PTB $\approx$31.51--31.52 in Table~\ref{tab:dp-mia-avg-comparison}). Therefore, the improved privacy of TTLoRA is not explained by a utility drop, but rather by reduced member/non-member separability in the calibrated loss space that the attacker exploits.

Overall, under identical DP budgets and the same attack protocol, TTLoRA provides a strictly improved privacy profile over LoRA while preserving essentially the same perplexity in this attack-evaluation regime.

\subsection{Non-Private Baseline: Utility and MIA Vulnerability}
\label{sec:nonprivatebaseline}
We establish a non-private baseline by evaluating standard fine-tuning and PEFT adaptation under the PreCurious MIA protocol. Starting from a pretrained GPT-2 checkpoint, we warm-start the reference model on the auxiliary split $D_{\mathrm{aux}}$ for 5 epochs and then attach a PEFT module (LoRA or TTLoRA) to obtain the adapted model. The adapted model is fine-tuned on $D_{\mathrm{train}}$ for up to 200 epochs with early stopping (patience = 5) to get the stable and converged model for privacy anaylsis. 

After convergence, we compute per-example losses on members ($D_{\mathrm{train}}$) and non-members ($D_{\mathrm{non}}$) and evaluate membership inference using the calibrated loss-gap score using equation~\ref{eq:precurious-ref-score}. Table~\ref{tab:nonprivate} summarizes average utility and attack metrics over even ranks $r\in\{2,4,\dots,16\}$ whereas Table \ref{tab:nonprivate-enron-only} details the rank-wise performance and privacy analysis for Enron dataset (detailed table for PTB in appendix Table \ref{tab:nonprivate-ptb-only}). Figure~\ref{fig:privacy-utility-nonprivate} visualizes the privacy--utility frontier. Complementarily, Figure~\ref{fig:loss-distribution} shows the calibrated loss distributions for members vs.\ non-members, where larger distributional gaps correspond to stronger membership signals and thus higher attack success. 

Our results indicate that \textbf{TTLoRA attains a distinctly improved privacy--efficiency trade-off compared to LoRA}, achieving comparable language-modeling utility while substantially reducing membership leakage with substantially fewer trainable parameters. On Enron at rank $r{=}2$, TTLoRA uses only 7.6K trainable parameters (vs.\ 110.6K for LoRA; $14.5\times$ fewer) and reaches a perplexity of 18.35 compared to 17.74 for LoRA, corresponding to only a 3.4\% relative difference in PPL (see Table \ref{tab:nonprivate-enron-only}). At this comparable utility, TTLoRA substantially decreases attack effectiveness, reducing attack ROC-AUC by 20.74 percentage points (92.23\% for LoRA vs. 71.49\% for TTLoRA) and reducing TPR at $\mathrm{FPR}{=}1\%$ by 21.14 percentage points (25.34\% for LoRA vs. 4.20\% for TTLoRA). Aggregated over even ranks $r\in\{2,4,\ldots,16\}$, TTLoRA achieves a lower mean attack AUC-ROC than LoRA (88.68\% vs.\ 94.91\%), i.e., a 6.2 percentage-point reduction. The low-FPR operating-point metric highlights an even larger privacy gain: $\mathrm{TPR}@\mathrm{FPR}{=}1\%$ decreases from 33.47\% (LoRA) to 16.19\% (TTLoRA), an absolute reduction of 17.28 percentage points. Consistent with these quantitative improvements, Figure~\ref{fig:loss-distribution} shows that TTLoRA yields substantially greater overlap between member and non-member calibrated loss distributions than LoRA, thereby weakening the separability signal exploited by membership inference attacks.

\begin{table}[h]
\centering
\caption{Non-private fine-tuning: Utility (PPL$\downarrow$) and MIA vulnerability (AUC$\downarrow$ and FPR$\downarrow$) averaged over even ranks (2-16) for LoRA and TTLoRA; $\downarrow$ means lower is better.}
\label{tab:nonprivate}
\small
\begin{tabular}{lccccc}
\toprule
\textbf{Method} & \textbf{Params} & \textbf{PPL$\downarrow$} & \textbf{AUC$\downarrow$} & \textbf{FPR@1\%$\downarrow$} \\
\midrule
\multicolumn{5}{c}{\textit{Enron Dataset}} \\
\midrule
FFT & 124.0M & 16.45 & 96.63\% & 26.24\% \\

LoRA (avg.) & 497.7K & \textbf{17.46} & 94.91\% & 33.47\% \\
TTLoRA (avg.) & \textbf{65.6K} & 18.05 & \textbf{88.68}\% & \textbf{16.19}\% \\
\midrule
\multicolumn{5}{c}{\textit{PTB Dataset}} \\
\midrule
FFT & 124.0M & 23.15 & 99.15\% & 64.94\% \\

LoRA (avg.) & 497.7K & \textbf{25.49} & 98.85\% & 32.51\% \\
TTLoRA (avg.) & \textbf{65.6K} & 26.66 & \textbf{97.17}\% & \textbf{29.67}\% \\
\bottomrule
\end{tabular}
\end{table}

\begin{figure}[h]
    \centering
    \includegraphics[width=\linewidth]{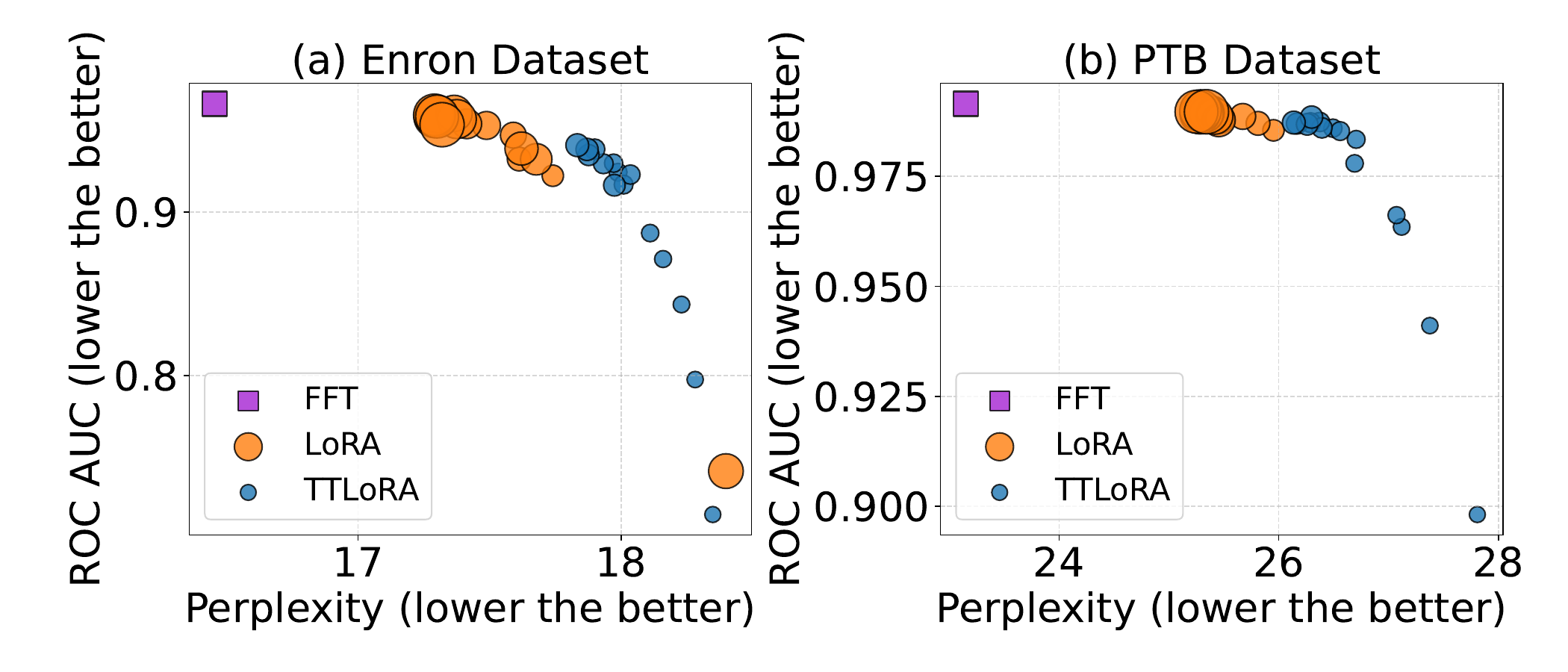}
    \caption{\textbf{Privacy-utility tradeoff in non-private fine-tuning.} Each point represents a rank configuration; marker size proportional to parameter count. TTLoRA achieves lower attack AUC (better privacy) with modest perplexity increase.}
    \label{fig:privacy-utility-nonprivate}
\end{figure}

\section{Discussion}
\label{sec:discussion}
Our results indicate that PEFT \emph{architecture} not just can lower the parameter count but also can substantially shape the privacy--utility trade-off in LLM adaptation. Across Enron and PTB, TTLoRA consistently reduces membership inference vulnerability relative to LoRA, and this advantage persists both without DP and under DP-SGD across the evaluated privacy budgets and TT-ranks. Importantly, these privacy gains are achieved in a markedly smaller adapter regime: TTLoRA operates with roughly 10K--150K trainable parameters, whereas LoRA requires roughly 100K--1,000K parameters over comparable rank ranges.
A key takeaway is that TTLoRA appears to suppress the membership signal exploited by calibrated loss-based attacks. The tensor-train parameterization restricts the update space through a chained, compositional structure, which can act as a stronger inductive bias than a single low-rank factorization in LoRA. Empirically, this manifests as greater overlap between member and non-member calibrated loss distributions and near-random-guess attack AUC under DP for TTLoRA, even as $\varepsilon$ increases. In contrast, LoRA’s leakage rises noticeably as the privacy budget relaxes, suggesting that additional capacity in a less constrained update space translates more directly into example-specific fitting.
Under DP-SGD, TTLoRA’s structure may provide a second benefit: improved robustness to optimization noise. Because the update is distributed across multiple small cores, each constrained by TT-ranks and sequential contractions, the effective degrees of freedom are reduced and gradients are implicitly regularized. This can make the learning dynamics less sensitive to the perturbations introduced by clipping and Gaussian noise, helping TTLoRA retain (and in some settings improve) perplexity relative to LoRA at the same $(\varepsilon,\delta)$.

\textbf{Practical implications.}
These findings suggest that privacy preserving adaptation need not rely solely on tighter privacy budgets or stronger noise multipliers. Instead, selecting a PEFT parameterization with an appropriate structural constraint can materially reduce empirical leakage while maintaining competitive utility, and can do so in an ultra-low-parameter regime that is attractive for deployment.

\textbf{Limitations.} Our evaluation focuses on GPT-2 (124M), two language modeling datasets, and a strong loss-based black-box MIA (PreCurious). While PreCurious captures an important and practical threat model, it does not cover the full spectrum of privacy risks (e.g., data extraction or broader memorization behaviors). In addition, our utility metric is perplexity; downstream task performance and instruction-following behaviors may exhibit different trade-offs.

\textbf{Future work.}
Several directions follow naturally from this study: (1) scaling TTLoRA-DP to larger modern LLMs (e.g., LLaMA-family and GPT-style models), where the adapter parameter savings and training-time efficiencies are even more consequential; (2) expanding evaluation beyond perplexity to downstream tasks and training regimes (instruction tuning, supervised fine-tuning, RLHF-style settings); (3) testing stronger and complementary privacy evaluations, including alternative MIAs, data extraction attacks, and memorization-oriented benchmarks; (4) combining TTLoRA-DP with orthogonal privacy techniques (e.g., federated learning or secure aggregation) to study end-to-end privacy in distributed settings; and (5) developing privacy accounting and algorithmic refinements that leverage TT structure—e.g., layer-wise clipping, structured noise, or tighter accounting under structural constraints—to further improve the privacy--utility frontier.

\section{Conclusion}
\label{sec:conclusion}
This work studies how PEFT architecture influences privacy leakage and utility under differentially private fine-tuning of language models. We establish TTLoRA as a structurally constrained PEFT approach and address the core systems challenge of enabling DP-SGD for TT-adapters by extending ghost clipping with cached contraction states, making exact per-example norm computation practical for TT-core backpropagation.

Empirically, TTLoRA improves the privacy--utility trade-off across two datasets and multiple privacy budgets. Under DP training, TTLoRA achieves consistently improved perplexity than LoRA at all evaluated $\varepsilon$ values while using substantially fewer trainable parameters.

Under the PreCurious membership inference attack protocol, TTLoRA stays close to the random-guess baseline across privacy budgets and ranks, whereas LoRA becomes increasingly vulnerable as the privacy budget relaxes. These privacy gains are not driven by degraded utility in the attack-evaluation pipeline.

Overall, our results suggest that TT decomposition is not only a compression mechanism but also a principled architectural bias for privacy-preserving LLM adaptation. TTLoRA provides a practical route to improving DP utility while reducing empirical membership leakage relative to widely used PEFT baselines under comparable privacy constraints.

\begin{acks}
This research was funded by the Los Alamos National Laboratory (LANL) Laboratory Directed Research and Development (LDRD) program under grants  20250850ECR and 20240868PRD3 and supported by LANL’s Institutional Computing Program, and by the U.S. Department of Energy National Nuclear Security Administration under Contract No. 89233218CNA000001. 
\end{acks}
\bibliographystyle{ACM-Reference-Format}
\bibliography{references}

\clearpage

\appendix

\section{Appendix A: Rank-wise Results and Extended Tables}
This appendix reports additional results that are omitted from the main paper due to space constraints. In the main paper, we primarily present \emph{rank-averaged} metrics to reduce sensitivity to any single adapter rank choice and to provide a robust comparison between LoRA and TTLoRA under differential privacy (DP). Here, we provide the \emph{full rank-wise breakdown} for (i) validation perplexity (PPL) across privacy budgets and (ii) membership inference attack (MIA) metrics for each rank and $\varepsilon$.

\paragraph{\textbf{How to read the tables}}
Across all appendix tables, lower $\downarrow$ is better for utility PPL and for attack metrics (AUC and FPR variants). Orange/blue highlights denote the best \emph{individual} configuration within LoRA/TTLoRA respectively, while bold black indicates the best \emph{average} between LoRA and TTLoRA (when an average row is present). Gray shading marks approximately parameter-matched configurations (to support controlled comparisons under comparable trainable parameter counts).

\paragraph{\textbf{Non-private rank-wise reference on PTB}}
Finally, Table~\ref{tab:nonprivate-ptb-only} provides non-private results on PTB across ranks. This table serves as a reference to interpret rank effects without DP noise and clipping. It also contextualizes the parameter--utility tradeoff across ranks: while larger ranks may yield diminishing improvements in PPL, they substantially increase trainable parameter counts, motivating the paper's focus on the parameter-efficient regime and the use of rank-averaged reporting in the main sections.

\paragraph{\textbf{Rank-wise DP perplexity trends}}
Table~\ref{tab:comprehensive_ppl_even} reports validation PPL for all even ranks $r\in\{2,4,\ldots,16\}$ across privacy budgets for Enron and PTB. These rank-wise results complement the rank-averaged trends in the main paper by showing that (1) PPL changes across ranks are typically modest within a fixed DP budget, and (2) the ordering across privacy budgets is consistent with the expected privacy--utility behavior (tighter privacy budgets generally yield higher PPL). This table also supports the interpretation that reporting rank-averaged PPL in the main paper is representative of the overall trend rather than dominated by a single rank.

\paragraph{\textbf{Rank-wise DP attack trends on Enron and PTB}}
Tables~\ref{tab:mia-dp-enron-only} and~\ref{tab:mia-dp-ptb-only} report rank-wise MIA results under DP for Enron and PTB respectively. For each $\varepsilon$, we report the attack AUC, FPR@10\%, and utility PPL. These tables make two points explicit:
(i) attack success generally increases as privacy becomes weaker (larger $\varepsilon$), consistent with standard DP expectations; and
(ii) TTLoRA exhibits systematically reduced attack success compared to LoRA on average, while maintaining comparable utility, aligning with the privacy--utility trends reported in the main paper.

\begin{table}[htpb]
  \centering
  \caption{\small Comparison of attack metrics (AUC, FPR@1\%, FPR@0.01\%) and utility (PPL) for LoRA, TTLoRA, and FFT on the PTB dataset. }
  \label{tab:nonprivate-ptb-only}
  \setlength{\tabcolsep}{3pt}
  \begin{tabular}{p{1.0cm} c c c c c c}
    \toprule
    Method & Rank & Params 
    & PPL 
    & AUC 
    & FPR@1\% 
    & FPR@0.01\% \\
    \midrule

    \rowcolor{gray!15}
    & 2  & 110.6k & \textbf{25.95} & \textbf{\textcolor{myorange}{98.55\%}} & 33.33\% & \textbf{\textcolor{myorange}{0.57\%}} \\
    & 4  & 221.2k & 25.67 & 98.87\% & 44.54\% & \textbf{\textcolor{myorange}{0.57\%}} \\
    & 6  & 331.8k & 25.46 & 98.81\% & 24.43\% & \textbf{\textcolor{myorange}{0.57\%}} \\
    & 8  & 442.4k & 25.45 & 98.78\% & \textbf{\textcolor{myorange}{23.56\%}} & \textbf{\textcolor{myorange}{0.57\%}} \\
    \textbf{LoRA}
    & 10 & 553.0k & 25.42 & 98.87\% & 27.30\% & \textbf{\textcolor{myorange}{0.57\%}} \\
    & 12 & 663.6k & 25.32 & 98.93\% & 28.45\% & \textbf{\textcolor{myorange}{0.57\%}} \\
    & 14 & 774.1k & \textbf{\textcolor{myorange}{25.29}} & 99.00\% & 39.08\% & \textbf{\textcolor{myorange}{0.57\%}} \\
    & 16 & 884.7k & 25.34 & 98.97\% & 39.37\% & \textbf{\textcolor{myorange}{0.57\%}} \\
    \cmidrule(lr){2-7}
    \textbf{Avg.} & -- & 497.7k & \textbf{25.49} & 98.85\% & 32.51\% & 0.57\% \\

    \midrule

    & 2  & 7.6k  & 27.81 & \textbf{\textcolor{myblue}{89.81\%}} & \textbf{\textcolor{myblue}{22.13\%}} & 1.15\% \\
    & 4  & 18.2k & 27.12 & 96.35\% & 39.66\% & \textbf{\textcolor{myblue}{0.57\%}} \\
    & 6  & 31.8k & 26.69 & 97.80\% & 22.99\% & \textbf{\textcolor{myblue}{0.57\%}} \\
    & 8  & 48.4k & 26.49 & 98.60\% & 34.20\% & \textbf{\textcolor{myblue}{0.57\%}} \\
    \textbf{TTLoRA}
    & 10 & 67.9k & 26.38 & 98.74\% & 54.60\% & 0.86\% \\
    \rowcolor{gray!15}
    & 12 & 90.4k & 26.39 & 98.61\% & 27.01\% & \textbf{\textcolor{myblue}{0.57\%}} \\
    \rowcolor{gray!10}
    & 14 & 115.9k & 26.26 & 98.69\% & 14.08\% & 0.57\% \\
    & 16 & 144.4k & \textbf{\textcolor{myblue}{26.14}} & 98.73\% & 22.70\% & \textbf{\textcolor{myblue}{0.57\%}} \\
    \cmidrule(lr){2-7}
    \textbf{Avg.} & -- & \textbf{65.6k} & 26.66 & \textbf{97.17\%} & \textbf{29.67\%} & \textbf{0.68\%} \\

    \midrule
    \textbf{FFT} & -- & 124.04M & 23.15 & 99.15\% & 64.94\% & 0.86\% \\
    \bottomrule
  \end{tabular}
\end{table}

%\clearpage

\begin{table*}[h]
\centering
\caption{Comparison of Perplexity (Lower $\downarrow$ the better) on Enron and PTB datasets across different privacy budgets ($\varepsilon$) for different ranks.}
\label{tab:comprehensive_ppl_even}
\small
\begin{tabular}{l l ccccc ccccc}
\toprule
\multirow{2}{*}{Method} & \multirow{2}{*}{Params} & \multicolumn{5}{c}{Enron ($\varepsilon$)} & \multicolumn{5}{c}{PTB ($\varepsilon$)} \\
\cmidrule(lr){3-7} \cmidrule(lr){8-12}
& & 0.5$\downarrow$ & 1.0$\downarrow$ & 3.0$\downarrow$ & 5.0$\downarrow$ & Non-Priv.$\downarrow$ & 0.5$\downarrow$ & 1.0$\downarrow$ & 3.0$\downarrow$ & 5.0$\downarrow$ & Non-Priv.$\downarrow$ \\
\midrule
\multicolumn{12}{l}{\textbf{LoRA}} \\
\rowcolor{gray!15} 
Rank 2 & 110.59K & \textbf{\textcolor{myorange}{27.95}} & \textbf{\textcolor{myorange}{26.88}} & \textbf{\textcolor{myorange}{25.65}} & \textbf{\textcolor{myorange}{25.27}} & 19.40 & \textbf{\textcolor{myorange}{59.13}} & \textbf{\textcolor{myorange}{56.04}} & \textbf{\textcolor{myorange}{51.60}} & \textbf{\textcolor{myorange}{49.74}} & 29.92 \\
Rank 4 & 221.18K & 28.38 & 27.11 & 25.98 & 25.64 & 18.95 & 59.54 & 57.14 & 53.19 & 51.70 & 29.22 \\
Rank 6 & 331.78K & 28.79 & 27.41 & 26.28 & 25.95 & 18.78 & 60.36 & 58.25 & 54.26 & 52.90 & 28.84 \\
Rank 8 & 442.37K & 29.05 & 27.68 & 26.58 & 26.25 & 18.73 & 60.32 & 58.68 & 54.89 & 53.56 & 28.64 \\
Rank 10 & 552.96K & 29.46 & 27.86 & 26.66 & 26.34 & 18.65 & 60.59 & 59.18 & 55.33 & 53.97 & 28.60 \\
Rank 12 & 663.55K & 29.89 & 28.09 & 26.81 & 26.47 & 18.63 & 60.65 & 59.62 & 55.84 & 54.42 & 28.61 \\
Rank 14 & 774.14K & 29.97 & 28.30 & 26.95 & 26.61 & \textbf{\textcolor{myorange}{18.58}} & 60.62 & 59.72 & 56.26 & 54.93 & 28.53 \\
Rank 16 & 884.74K & 30.84 & 28.51 & 28.18 & 27.96 & \textbf{\textcolor{myorange}{18.58}} & 60.76 & 60.10 & 56.64 & 55.19 & \textbf{\textcolor{myorange}{28.51}} \\
\midrule
Avg. & 497.7K & 29.29 & 27.62 & 26.64 & 26.44 & \textbf{18.79} & 60.41 & 58.59 & 54.75 & 53.81 & \textbf{28.86} \\
\midrule
\multicolumn{12}{l}{\textbf{TTLoRA}} \\
Rank 2 & 7.63K & 27.71 & 26.51 & 24.69 & \textbf{\textcolor{myblue}{24.34}} & 22.25 & \textbf{\textcolor{myblue}{57.56}} & \textbf{\textcolor{myblue}{54.03}} & \textbf{\textcolor{myblue}{49.66}} & 48.25 & 37.32 \\
Rank 4 & 18.24K & \textbf{\textcolor{myblue}{27.28}} & \textbf{\textcolor{myblue}{25.92}} & \textbf{\textcolor{myblue}{24.65}} & 24.40 & 21.08 & 58.25 & 54.71 & 49.59 & \textbf{\textcolor{myblue}{47.59}} & 33.58 \\
Rank 6 & 31.82K & 27.62 & 26.52 & 24.92 & 24.51 & 20.42 & 57.60 & 54.89 & 50.35 & 48.45 & 32.04 \\
Rank 8 & 48.38K & 27.63 & 26.90 & 25.57 & 25.11 & 19.97 & 58.05 & 55.77 & 51.82 & 50.03 & 30.84 \\
Rank 10 & 67.92K & 27.29 & 26.44 & 25.20 & 24.75 & 19.65 & 58.58 & 55.47 & 50.01 & 48.00 & 30.09 \\
\rowcolor{gray!15} 
Rank 12 & 90.43K & 27.47 & 26.74 & 25.92 & 25.45 & 19.40 & 58.92 & 55.91 & 50.54 & 48.47 & 29.58 \\
\rowcolor{gray!10} 
Rank 14 & 115.92K & 27.44 & 26.67 & 25.94 & 25.58 & 19.17 & 58.77 & 55.60 & 50.96 & 48.96 & 29.25 \\
Rank 16 & 144.38K & 27.84 & 27.04 & 26.27 & 25.93 & \textbf{\textcolor{myblue}{19.02}} & 60.02 & 56.77 & 51.96 & 49.91 & \textbf{\textcolor{myblue}{28.91}} \\
\midrule
\textbf{Avg.} & \textbf{65.6K} & \textbf{27.52} & \textbf{26.55} & \textbf{25.37} & \textbf{24.99} & 20.14 & \textbf{58.47} & \textbf{55.39} & \textbf{50.61} & \textbf{48.71} & 31.45 \\
\midrule
FFT & 124.04M & 28.81 & 27.05 & 25.38 & 24.92 & 14.31 & 56.18 & 53.38 & 49.23 & 47.79 & 19.75 \\
\bottomrule
\end{tabular}
\end{table*}

\begin{table*}[htbp]
  \centering
  \caption{Comparison of Attack Metrics (AUC and FPR@10\%) and utility (PPL) for LoRA and TTLoRA across Enron dataset for $\varepsilon$ values. Lower $\downarrow$ is better for AUC, FPR@10\%, and PPL.}
  \label{tab:mia-dp-enron-only}
  \small
\begin{tabular}{c | c c c | c c c | c c c | c c c}
    \toprule
    \textbf{Method} & \multicolumn{12}{c}{\textbf{Epsilon} ($\varepsilon$)} \\
    \cmidrule(lr){2-13}
    & \multicolumn{3}{c}{0.5} & \multicolumn{3}{c}{1.0} & \multicolumn{3}{c}{3.0} & \multicolumn{3}{c}{5.0}  \\
    \cmidrule(lr){2-4} \cmidrule(lr){5-7} \cmidrule(lr){8-10} \cmidrule(lr){11-13}
    & \textbf{AUC} & \textbf{PPL} & \textbf{FPR@10\%} & \textbf{AUC} & \textbf{PPL} & \textbf{FPR@10\%} & \textbf{AUC} & \textbf{PPL} & \textbf{FPR@10\%} & \textbf{AUC} & \textbf{PPL} & \textbf{FPR@10\%}  \\
    \midrule
    \textbf{LoRA} &&&&&&&&&&&&\\
    \rowcolor{gray!15} 
    Rank2  & $54.79\%$ & $20.72$ & $11.69\%$ & $56.91\%$ & $20.72$ & $12.89\%$ & $61.14\%$ & $20.71$ & $17.24\%$ & $62.32\%$ & $20.70$ & $19.49\%$ \\
    Rank4  & $54.33\%$ & $20.72$ & $12.59\%$ & $56.44\%$ & $20.72$ & $14.69\%$ & $60.57\%$ & $20.71$ & $17.00\%$ & $61.99\%$ & $20.70$ & $18.14\%$ \\
    Rank6  & $52.25\%$ & $20.72$ & $12.74\%$ & $54.11\%$ & $20.72$ & $13.19\%$ & $57.86\%$ & $20.71$ & $17.84\%$ & $59.36\%$ & $20.71$ & $16.34\%$ \\
    Rank8  & \textbf{\textcolor{myorange}{50.06\%}} & $20.72$ & $9.90\%$  & $50.79\%$ & $20.72$ & $10.49\%$ & $52.76\%$ & $20.72$ & $11.99\%$ & \textbf{\textcolor{myorange}{53.94\%}} & $20.72$ & \textbf{\textcolor{myorange}{11.69\%}} \\
    Rank10 & $51.14\%$ & $20.72$ & $10.04\%$ & $53.08\%$ & $20.72$ & $11.69\%$ & $57.30\%$ & $20.71$ & $15.89\%$ & $58.94\%$ & $20.71$ & $16.79\%$ \\
    Rank12 & $51.91\%$ & $20.71$ & $11.00\%$ & \textbf{\textcolor{myorange}{50.72\%}} & $20.71$ & $10.94\%$ & \textbf{\textcolor{myorange}{52.67\%}} & $20.71$ & \textbf{\textcolor{myorange}{11.84\%}} & $54.44\%$ & $20.71$ & $13.34\%$ \\
    Rank14 & $53.27\%$ & $20.71$ & $11.39\%$ & $54.71\%$ & $20.71$ & $12.89\%$ & $58.25\%$ & $20.71$ & $13.49\%$ & $59.74\%$ & $20.71$ & $16.64\%$ \\
    Rank16 & $52.15\%$ & $20.72$ & \textbf{\textcolor{myorange}{9.75\%}}  & $52.99\%$ & $20.72$ & \textbf{\textcolor{myorange}{10.04\%}} & $55.36\%$ & $20.72$ & $13.94\%$ & $56.72\%$ & $20.72$ & $14.69\%$ \\
    \midrule
    \textbf{Avg.} & $52.49\%$ & $20.72$ & $11.14\%$ & $53.72\%$ & $20.72$ & $12.10\%$ & $56.99\%$ & $20.71$ & $14.90\%$ & $58.43\%$ & $20.71$ & $15.89\%$ \\
    \midrule
    \textbf{TTLoRA} &&&&&&&&&&&&\\
    Rank2  & $52.87\%$ & $20.78$ & $12.44\%$ & $53.65\%$ & $20.75$ & $13.19\%$ & $55.54\%$ & $20.73$ & $14.39\%$ & $56.46\%$ & $20.72$ & $14.54\%$ \\
    Rank4  & $52.37\%$ & $20.73$ & $9.45\%$  & $52.12\%$ & $20.72$ & $9.15\%$  & $51.27\%$ & $20.72$ & $9.00\%$  & $50.84\%$ & $20.72$ & $9.00\%$  \\
    Rank6  & $50.75\%$ & $20.72$ & $10.64\%$ & $51.05\%$ & $20.72$ & $10.94\%$ & $51.79\%$ & $20.72$ & $11.39\%$ & $52.19\%$ & $20.72$ & $11.09\%$ \\
    Rank8  & $50.77\%$ & $20.71$ & $9.00\%$  & $51.13\%$ & $20.71$ & $9.00\%$  & $51.88\%$ & $20.71$ & $9.90\%$  & $52.47\%$ & $20.71$ & $10.04\%$ \\
    Rank10 & $50.74\%$ & $20.72$ & $13.49\%$ & $50.94\%$ & $20.72$ & $13.64\%$ & $51.27\%$ & $20.72$ & $12.74\%$ & $51.51\%$ & $20.72$ & $13.04\%$ \\
    \rowcolor{gray!15} 
    Rank12 & $51.10\%$ & $20.72$ & $11.54\%$ & $50.67\%$ & $20.72$ & $12.44\%$ & $50.23\%$ & $20.72$ & $13.19\%$ & $50.63\%$ & $20.72$ & $13.49\%$ \\
    \rowcolor{gray!10} 
    Rank14 & \textbf{\textcolor{myblue}{50.39\%}} & $20.72$ & \textbf{\textcolor{myblue}{8.40\%}}  & \textbf{\textcolor{myblue}{50.27\%}} & $20.72$ & \textbf{\textcolor{myblue}{8.25\%}}  & \textbf{\textcolor{myblue}{50.06\%}} & $20.72$ & \textbf{\textcolor{myblue}{8.10\%}}  & \textbf{\textcolor{myblue}{50.07\%}} & $20.72$ & \textbf{\textcolor{myblue}{7.80\%}}  \\
    Rank16 & $51.91\%$ & $20.72$ & $10.34\%$ & $51.97\%$ & $20.72$ & $10.34\%$ & $52.05\%$ & $20.72$ & $10.79\%$ & $52.10\%$ & $20.72$ & $10.79\%$ \\
    \midrule
    \textbf{Avg.} & $\textbf{51.36\%}$ & $20.72$ & $\textbf{10.66\%}$ & $\textbf{51.48\%}$ & $20.72$ & $\textbf{10.87\%}$ & $\textbf{51.76\%}$ & $20.72$ & $\textbf{11.19\%}$ & $\textbf{52.03\%}$ & $20.72$ & $\textbf{11.22\%}$ \\
    \midrule 
    \textbf{FFT} & 54.38\% & 20.96 & 13.04\% & 55.22\% & 20.88 & 13.04\% & 56.85\% & 20.76 & 13.34\% & 57.72\% & 20.72 & 13.49\%\\
    \bottomrule
\end{tabular}
\end{table*}

\clearpage

\begin{table*}[htbp]
  \centering
  \caption{Comparison of Attack Metrics (AUC and FPR@10\%) and utility (PPL) for LoRA and TTLoRA across PTB dataset for $\varepsilon$ values. Lower $\downarrow$ is better for AUC, FPR@10\%, and PPL.}
  \label{tab:mia-dp-ptb-only}
  \begin{tabular}{c | c c c | c c c | c c c | c c c}
    \toprule
    \textbf{Method} & \multicolumn{12}{c}{\textbf{Epsilon} ($\varepsilon$)} \\
    \cmidrule(lr){2-13}
    & \multicolumn{3}{c}{0.5} & \multicolumn{3}{c}{1.0} & \multicolumn{3}{c}{3.0} & \multicolumn{3}{c}{5.0}  \\
    \cmidrule(lr){2-4} \cmidrule(lr){5-7} \cmidrule(lr){8-10} \cmidrule(lr){11-13}
    & \textbf{AUC\%} & \textbf{PPL} & \textbf{FPR@10\%} & \textbf{AUC\%} & \textbf{PPL} & \textbf{FPR@10\%} & \textbf{AUC\%} & \textbf{PPL} & \textbf{FPR@10\%} & \textbf{AUC\%} & \textbf{PPL} & \textbf{FPR@10\%}  \\
    \midrule
    \textbf{LoRA}&&&&&&&&&&&&\\
    \rowcolor{gray!15}
    Rank2  & 53.90\% & 32.32 & 11.21\% & 55.02\% & 31.91 & 12.07\% & 58.41\% & 31.42 & 16.67\% & 60.18\% & 31.25 & 18.39\%  \\
    Rank4  & \textbf{\textcolor{myorange}{51.35\%}} & 32.31 & \textbf{\textcolor{myorange}{7.76\%}}  & 52.69\% & 31.90 & 10.06\% & 56.75\% & 31.40 & \textbf{\textcolor{myorange}{11.49\%}} & 58.88\% & 31.23 & 16.38\%  \\
    Rank6  & 52.14\% & 32.35 & 13.22\% & 53.53\% & 31.94 & 15.23\% & 57.47\% & 31.43 & 19.83\% & 59.78\% & 31.27 & 21.26\%  \\
    Rank8  & 52.89\% & 32.34 & 10.92\% & \textbf{\textcolor{myorange}{51.96\%}} & 31.93 & 12.07\% & \textbf{\textcolor{myorange}{50.77\%}} & 31.42 & \textbf{\textcolor{myorange}{11.49\%}} & \textbf{\textcolor{myorange}{52.66\%}} & 31.26 & \textbf{\textcolor{myorange}{13.51\%}}  \\
    Rank10 & 53.02\% & 32.36 & 12.93\% & 54.22\% & 31.95 & 15.23\% & 57.70\% & 31.45 & 19.25\% & 59.70\% & 31.29 & 22.41\%  \\
    Rank12 & 52.30\% & 32.32 & 8.91\%  & 53.37\% & 31.91 & 9.77\%  & 56.29\% & 31.41 & 12.36\% & 58.13\% & 31.25 & 15.80\%  \\
    Rank14 & 53.88\% & 32.38 & 13.51\% & 55.17\% & 31.97 & 13.79\% & 58.96\% & 31.46 & 18.68\% & 61.16\% & 31.31 & 21.84\%  \\
    Rank16 & 51.49\% & 32.34 & 8.62\%  & 52.61\% & 31.93 & \textbf{\textcolor{myorange}{9.20\%}}  & 56.06\% & 31.43 & 12.36\% & 58.11\% & 31.26 & 15.52\%  \\
    \midrule
    \textbf{Avg.} & 52.62\% & 32.34 & 10.88\% & 53.57\% & 31.93 & 12.18\% & 56.55\% & 31.43 & 15.27\% & 58.58\% & 31.27 &18.14\%  \\
    \midrule
    \textbf{TTLoRA}&&&&&&&&&&&&\\
    Rank2  & 52.00\% & 32.31 & 8.91\%  & 51.55\% & 31.91 & 8.62\%  & \textbf{\textcolor{myblue}{50.03\%}} & 31.41 & \textbf{\textcolor{myblue}{6.90\%}}  & 51.05\% & 31.25 & 6.90\%  \\
    Rank4  & 52.52\% & 32.32 & 8.05\%  & 52.97\% & 31.91 & 7.18\%  & 53.55\% & 31.42 & 8.62\%  & 53.87\% & 31.25 & 10.63\% \\
    Rank6  & 52.98\% & 32.32 & \textbf{\textcolor{myblue}{6.03\%}}  & 52.91\% & 31.92 & \textbf{\textcolor{myblue}{6.32\%}}  & 52.70\% & 31.43 & 7.76\%  & 52.49\% & 31.26 & \textbf{\textcolor{myblue}{6.61\%}}  \\
    Rank8  & 50.83\% & 32.35 & 12.07\% & 51.05\% & 31.94 & 11.78\% & 51.75\% & 31.45 & 11.21\% & 52.17\% & 31.28 & 10.34\% \\
    Rank10 & 50.90\% & 32.36 & 12.07\% & 50.81\% & 31.96 & 12.07\% & 50.67\% & 31.46 & 12.36\% & \textbf{\textcolor{myblue}{50.58\%}} & 31.30 & 12.93\% \\
    \rowcolor{gray!15}
    Rank12 & 52.03\% & 32.35 & 9.48\%  & 52.31\% & 31.95 & 9.77\%  & 53.24\% & 31.45 & 9.48\%  & 53.89\% & 31.28 & 10.92\% \\
    \rowcolor{gray!10}
    Rank14 & 51.46\% & 32.35 & 8.62\%  & 51.58\% & 31.95 & 8.62\%  & 51.96\% & 31.44 & 8.05\%  & 52.26\% & 31.28 & 8.05\%  \\
    Rank16 & \textbf{\textcolor{myblue}{50.25\%}} & 32.38 & 10.06\% & \textbf{\textcolor{myblue}{50.02\%}} & 31.98 & 10.34\% & 50.66\% & 31.48 & 11.21\% & 51.18\% & 31.31 & 12.36\% \\
    \midrule
    \textbf{Avg.} & \textbf{51.62\%} & 32.34 & \textbf{9.41\%} & \textbf{51.65\%} & 31.94& \textbf{9.34\%} & \textbf{51.82\%} & 31.44& \textbf{9.45\%} & \textbf{52.19\%} & 31.28 & \textbf{9.84\%}  \\
    \midrule 
    \textbf{FFT} & 51.21\% & 31.74 & 8.33\% & 50.94\% & 31.70 & 8.33\% & 50.69\% & 31.61 & 13.79\% & 52.23\% & 31.57 & 14.37\% \\
    \bottomrule
  \end{tabular}
\end{table*}

\end{document}